\documentclass{elsarticle}

\usepackage{fullpage}
\usepackage{epsfig}

\newcommand{\noop}[1][1]{}
\usepackage{amsmath,amssymb,epsfig}
\usepackage{latexsym}
\usepackage{hyperref}
\usepackage{mathrsfs}
\usepackage{enumerate}
\usepackage{pbox}

\newtheorem{mydef}{Definition}

\newtheorem{mythm}{Theorem}
\newtheorem{mylem}{Lemma}
\newtheorem{mycor}{Corollary}

\newcounter{thm} 

\newcommand{\bb}{\mathbf}
\newcommand{\dg}{{+}} 
\newlength{\digitwidth}

\newcommand{\ignore}[1]{}

\begin{document}

%
%

\title{Incremental Computation of Pseudo-Inverse of Laplacian\\ {\small \em Theory and Applications}}
\author{Gyan Ranjan\corref{cor1}}
\ead{granjan@cs.umn.edu}
\author{Zhi-Li Zhang}
\ead{zhzhang@cs.umn.edu} 
\author{Daniel Boley}
\ead{boley@cs.umn.edu}
\address{Dept. of Computer Science, University of Minnesota, Twin Cities, USA.}
\cortext[cor1]{Corresponding author.}
%

%
%

\begin{abstract}
A divide-and-conquer based approach for computing the Moore-Penrose pseudo-inverse of the 
combinatorial Laplacian matrix $(\bb L^+)$ of a simple, undirected graph is proposed. 
The nature of the underlying sub-problems is studied in detail by means of an elegant interplay 
between $\bb L^+$ and the effective resistance distance $(\Omega)$.   
Closed forms are provided for a novel {\em two-stage} process that helps compute the 
pseudo-inverse incrementally.  
Analogous scalar forms are obtained for the converse case, that of structural regress, 
which entails the breaking up of a graph into disjoint components 
through successive edge deletions.     
The scalar forms in both cases, show absolute element-wise independence at all stages, 
thus suggesting potential parallelizability. 
Analytical and experimental results are presented for dynamic (time-evolving) graphs as well 
as large graphs in general (representing real-world networks). 
An order of magnitude reduction in computational time is achieved for dynamic graphs; 
while in the general case, our approach performs better in practice than 
the standard methods, even though the worst case theoretical complexities may remain the same:  
an important contribution with consequences  to the study of online social networks.     
\end{abstract}

%
%

\maketitle

%
%

\section{Introduction}
\label{Sec:Introduction}
The combinatorial Laplacian matrix of a graph finds use in various aspects of structural analysis
\cite{Alon86, Chung97, DonathHoffman, Fiedler73, Fiedler75, Fouss07,
GobelJagers74, Mohar91, Mohar92,  Mohar97, Nilli91, ShiMalik97, Luxburg07}. 
The eigen spectrum of the Laplacian determines significant topological characteristics of the graph,  
such as minimal cuts, clustering and the number of spanning trees  
\cite{Biggs93, Fiedler73, Fiedler75, Luxburg07}.      
Likewise, the Moore-Penrose pseudo-inverse and the sub-matrix inverses of the Laplacian 
have evoked great interest in recent times. Their applications span fields as diverse as 
probability and mathematical chemistry, collaborative recommendation systems and 
social networks, epidemiology and infrastructure planning 
\cite{Fouss07,  IsaacsonMadsen76, Kirkland97,  KleinRandic93, Newman05, RanjanZhang11a, Xiao03}.  
A brief discussion of the specific applications is provided for reference in a subsequent section  
(c.f. \textsection\ref{Sec:RelWork}). 
Alas, despite such versatility, the pseudo-inverse and the sub-matrix inverses of the Laplacian 
suffer a practical handicap. These matrices are notoriously expensive to compute. 
The standard matrix factorization and inversion based methods    
employed to compute them \cite{Ben-IsraelGreville03, Xiao03}, incur an $O(n^3)$ computational time, 
$n$ being the order of the graph (number of vertices in the graph). 
This clearly impedes their utility particularly when the graphs are either dynamic, i.e. changing with time, 
or simply of large orders, i.e. have millions of nodes. Online social networks $(OSN)$, 
typically represented as graphs, qualify on both counts.  
With time, the number of users as well as the relationships between them changes, thus requiring 
regular re-computations. 
As for size, a popular OSN, such as Facebook and Youtube, may easily have hundreds of millions 
of users.  An $O(n^3)$ cost, therefore, is clearly undesirable and an approach for incremental updates 
is imperative, particularly given that such changes, in most cases, may be local in nature.  

In this work, we provide a novel divide-and-conquer based approach for computing the 
Moore-Penrose pseudo-inverse of the Laplacian for an undirected graph which, in turn, 
determines all of its sub-matrix inverses as well. 
The divide operation in our approach entails determining an arbitrary {\em connected bi-partition} 
of the graph $G(V, E)$ --- a cut of the graph that is made up of exactly two 
connected sub-graphs (say $G_1(V_1, E_1)$ and $G_2(V_2, E_2)$)
--- by deleting $\kappa$ edges from it.   
As $G_1(V_1, E_1)$ and $G_2(V_2, E_2)$ are simple and connected themselves, 
the pseudo-inverse of their Laplacians, when computed, constitute solutions to two independent 
sub-problems. Better still, they can be computed in parallel (given two machines instead of one). 
In the conquer step, we recombine these solutions in an iterative manner by re-introducing the edges 
in the cut set one at a time to reconstruct the original graph and obtain the overall pseudo-inverse in 
the process. Note that, this process yields a sequence of intermediate spanning sub-graphs of $G$,  
(say $\{G_1, G_2 \} \rightarrow G_3 \rightarrow G_4 \rightarrow ... \rightarrow G_{\kappa+2}$), 
where $G_{\kappa+2} = G(V, E)$. The first transition $\{G_1, G_2 \} \rightarrow G_3$ represents a point 
of {\em singularity} in our method whence the disjoint components $\{G_1, G_2\}$ get connected 
through a {\em bridge} edge to yield $G_3$, a sub-graph with exactly one component. 
We call this stage the {\em first join} in our process. 
Post the first join, all intermediate sub-graphs from $G_4$ to $G_{\kappa+2}$  
are obtained by introducing an edge in a sub-graph that is already connected.  
We call this {\em edge firing} (details in a subsequent section).  
We then show that the pseudo-inverse of the Laplacian for any intermediate sub-graph in this sequence 
is determined entirely in terms of the pseudo-inverse of the Laplacian for its predecessor. 
Our results, presented in an element-wise scalar form, reveal several interesting properties 
of the sub-problems. 
First and foremost, if $n$ be the order of the graph $G(V, E)$, then the cost incurred at each 
intermediate stage is $O(n^2)$ if the solution to the sub-problems for the immediate predecessor 
is known.
Therefore, the cost of computing the pseudo-inverse for $G(V, E)$ is $O(\kappa \cdot n^2)$, 
if the pseudo-inverses for $G_1(V_1, E_1)$ and $G_2(V_2, E_2)$ are known.   
Secondly, using these forms, each element of the pseudo-inverse for an intermediate graph can 
be computed independent of the other elements. Hence, given multiple machines, the overall 
computational time is reduced further through parallelization.    
Moreover, we obtain similar closed form solutions for the case of structural regress 
of the graph, i.e. when vertices or edges are deleted from it.   
A straightforward consequence is that the pseudo-inverses for dynamic time-evolving graphs, 
such as OSNs, can now be updated when a node joins or leaves the 
network or an edge (a relationship) appears/disappears in it,  at an $O(n^2)$ cost overall 
(as $\kappa << n$).  

Last but not least, we use these insights to compute the pseudo-inverses of the Laplacians of 
large real-world networks from the domain of online social networks.   
Real-world networks, and social ones in particular, are reported to have some notable characteristics 
such as edge sparsity, {\em power-law} and scale-free degree distributions 
\cite{Barabasi99, Barabasi00, Faloutsos99}, {\em small-world} characteristics \cite{Watts98} etc.
Given these properties, we note that interesting algorithms (heuristics) can be developed for fast 
and parallel computations for the general case based on our divide-and-conquer strategy. 
Thus, even though the theoretical worst  case costs stay at $O(n^3)$ for general graphs, the practical 
gains are significant enough to warrant attention. We discuss both analytical and experimental 
aspects of these in detail in the subsequent sections.    

The rest of the paper is organized into the following sections: 
we begin by introducing the preliminaries of our work --- the pseudo-inverse and the 
sub-matrix inverses of the Laplacian along with their properties;  
and the interplay of the pseudo-inverse and the effective resistance distance ---  
in \textsection\ref{Sec:LapSubInvDist}. 
In \textsection\ref{Sec:TwoToOne}, we describe our divide-and-conquer strategy involving 
connected bi-partitions and the {\em two-stage} process for computing 
the Moore-Penrose pseudo-inverse of the Laplacian. Relevant scalar forms are presented 
in each case.   
In \textsection\ref{Sec:OneToTwo}, we establish the same closed forms for a graph in regress i.e.  
deleting edges one at a time until the graph breaks into two.  
We then apply the divide-and-conquer methodology to compute the pseudo-inverses 
for dynamically changing graphs as well as those of real world networks in 
\textsection\ref{Sec:CompAndParallel}.  
In \textsection\ref{Sec:RelWork} we briefly overview related literature discussing specific application 
scenarios.  
The paper is finally concluded in \textsection\ref{Sec:Conclusion} with a summary of results and 
a discussion of potential future works. 
\
\

%
%

%
%

\section{The Laplacian, Sub-Matrix Inverses and A Distance Function}
\label{Sec:LapSubInvDist}
In this section, we provide a brief introduction to the set of matrices studied in this work, 
namely, the combinatorial Laplacian of a graph ($\bb L$), its Moore-Penrose pseudo-inverse 
($\bb L^+$) and the set of sub-matrix inverses of $\bb L$ (\textsection\ref{SubSec:LapPInv}).
We then demonstrate how all the 
sub-matrix inverses of the Laplacian can be computed in terms of the pseudo-inverse 
in \textsection\ref{SubSec:SubInvOfLap}. 
Finally,  in \textsection\ref{SubSec:ERDistLp} we describe the relationship between the 
{\em effective resistance distance}, a Euclidean metric, and the elements of the Moore-Penrose 
pseudo-inverse of the Laplacian --- an equivalence that we exploit to great advantage in the 
rest of this work. 
\subsection{The Laplacian and its Moore-Penrose Pseudo-Inverse}
\label{SubSec:LapPInv}
Let $G(V, E)$ be a simple, connected and undirected graph. 
We denote by $n = |V(G)|$ the number of nodes/vertices in $G$, 
also called the {\em order} of the graph $G$, and by $m = |E(G)|$ the number of links/edges.  
The adjacency matrix of $G(V, E)$ is defined as $\bb A \in \Re^{n \times n}$, 
with elements $[\bb A]_{xy} = a_{xy} = a_{yx} =  [\bb A]_{yx} = w_{ij}$, if $x \neq y$ and 
$e_{xy} \in E(G)$ is an edge; $0$ otherwise. Here, the weight of the edge $w_{ij}$ is a 
measure of {\em affinity} between nodes $i$ and $j$. Clearly, $\bb A$ is real and symmetric. 
The degree matrix $\bb D$, is a diagonal matrix where $[\bb D]_{xx} = d_{xx} = d(x) = \sum_{y \in V(G)} a_{xy}$,  
is the weighted degree of node $x \in V(G)$; the sum of all edge weights (affinities) emanating from $x$. 
Also, $vol(G) = \sum_{x \in V(G)} d(x)$, is called the {\em volume} of the graph $G$ --- the sum total of 
affinities between all pairs of vertices in $G$. 
The combinatorial Laplacian of the graph is then given by: 
\begin{equation}
\label{equ:Laplacian}
\bb L = \bb D - \bb A
\end{equation}
It is easy to see, from the definition in $(\ref{equ:Laplacian})$ above, that the 
Laplacian $\bb L$ is a real, symmetric and doubly-centered matrix (each 
row/column sum is 0). More importantly, $\bb L$ admits an eigen decomposition of the form 
$\bb L = \bb \Phi \bb \Lambda \bb \Phi'$ where the columns of $\bb \Phi$ constitute 
the set of orthogonal eigen vectors of $\bb L$ and $\bb \Lambda$ is a diagonal 
matrix with $[\Lambda]_{ii} = \lambda_{i}: 1 \leq i \leq n$; being the $n$ eigen values of $\bb L$. 
It is well established that for a connected undirected graph $G(V, E)$, $\bb L$ is positive 
semi-definite i.e. it has a unique smallest eigen value $\lambda_1 = 0$. 
The rest of the $n - 1$ eigen values are all positive.    
Thus, $\bb L$ is rank deficient ($rank(\bb L) = n - 1 < n$) and consequently singular. 
Its inverse, in the usual sense, does not exist.      

However, the Moore-Penrose pseudo-inverse of $\bb L$, denoted henceforth by 
$\bb L^+$ does exist and is unique. Following constitute the basic 
properties of  $\bb L^+$ as a unique generalized inverse of $\bb L$ \cite{Ben-IsraelGreville03}:  
\begin{equation}
a.~~~ \bb L \bb L^+ \bb L = \bb L ~~~~~~~~~~~~ b.~~~ \bb L^+ \bb L \bb L^+ = \bb L^+
~~~~~~~~~~~~
c.~~~ (\bb L \bb L^+)' = \bb L \bb L^+ ~~~~~~~ d.~~~(\bb L^+ \bb L)' = \bb L^+ \bb L
\end{equation}
Like $\bb L$, $\bb L^+$ is also real, symmetric, doubly centered 
and positive semi-definite. Moreover, the eigen decomposition of $\bb L^+$ is given 
by $\bb L^+ = \Phi \Lambda^+ \Phi'$, with the same set of orthogonal eigen-vectors 
as that of $\bb L$. The set of eigen values of $\bb L^+$, given by the diagonal of the 
matrix $\bb \Lambda^+$, is composed of $\lambda^+_1 = 0$ and the reciprocals of the 
positive eigen-values of $\bb L$. 
We denote by $l^+_{xy}$, the element in the $x^{th}$ row and $y^{th}$ column of $\bb L^+$ 
(a convention followed for all matrices henceforth). 
We emphasize that even when the matrix $\bb L$ is sparse (which is the case with real world networks), 
$\bb L^+$ is always a full matrix. In fact, for a connected graph, all the elements of $\bb L^+$ are non-zero.      

A straightforward approach for computing $\bb L^+$ is through 
the eigen-decomposition of $\bb L$, followed by an inversion of its non-zero eigen values, 
and finally reassembling the matrix as discussed above. 
In practice, however, mathematical software, such as MATLAB, use singular value 
decomposition to compute the pseudo-inverse of matrices (c.f. $pinv$ in the standard library). 
This general SVD based method does not exploit the special structural properties of 
$\bb L$ and incurs $O(n^3)$ computational time, $n$ being the number of nodes 
in the graph.    
An alternative has recently been proposed in \cite{Xiao03} specifically for computing 
$\bb L^+$ for a simple, connected, undirected graph. 
A $rank(1)$ perturbation of the matrix $\bb L$ makes it invertible. 
$\bb L^+$ can then be computed from this perturbed matrix as follows: 
\begin{equation}
\label{equ:PertInvLp}
\bb L^+ = \left(\bb L + \frac{1}{n} \bb J\right)^{-1} - \frac{1}{n}\bb J 
\end{equation}
\noindent where $\bb J \in \Re^{n \times n}$ is a matrix of all $1's$. 
Although the theoretical cost for this method is also $O(n^3)$, in practice it 
works faster for graphs of arbitrary orders and edge densities than the standard 
$pinv$ method. But the proof of this pudding is in computing!   
So, to put into context the notion, we present a numerical analysis  
over Erd\"{o}s-R\'{e}nyi graphs (ER-graphs) of varying orders and edge densities.  
An ER-graph is a random graph determined by parameters $(n, \rho)$, where 
$n$ is the order of the graph and $\rho$ is the uniform probability for the occurrence 
of any arbitrary (undirected) edge in the graph \cite{Bollobas01}. 
We use a dedicated machine with a quad-core AMD Opteron processor 
$(800 ~Mhz/core)$ and $48 ~GB$ of primary memory.  

\begin{figure}[t]
\centerline{\begin{tabular}{cc}
\scalebox{0.8}{\includegraphics[width = 100mm]{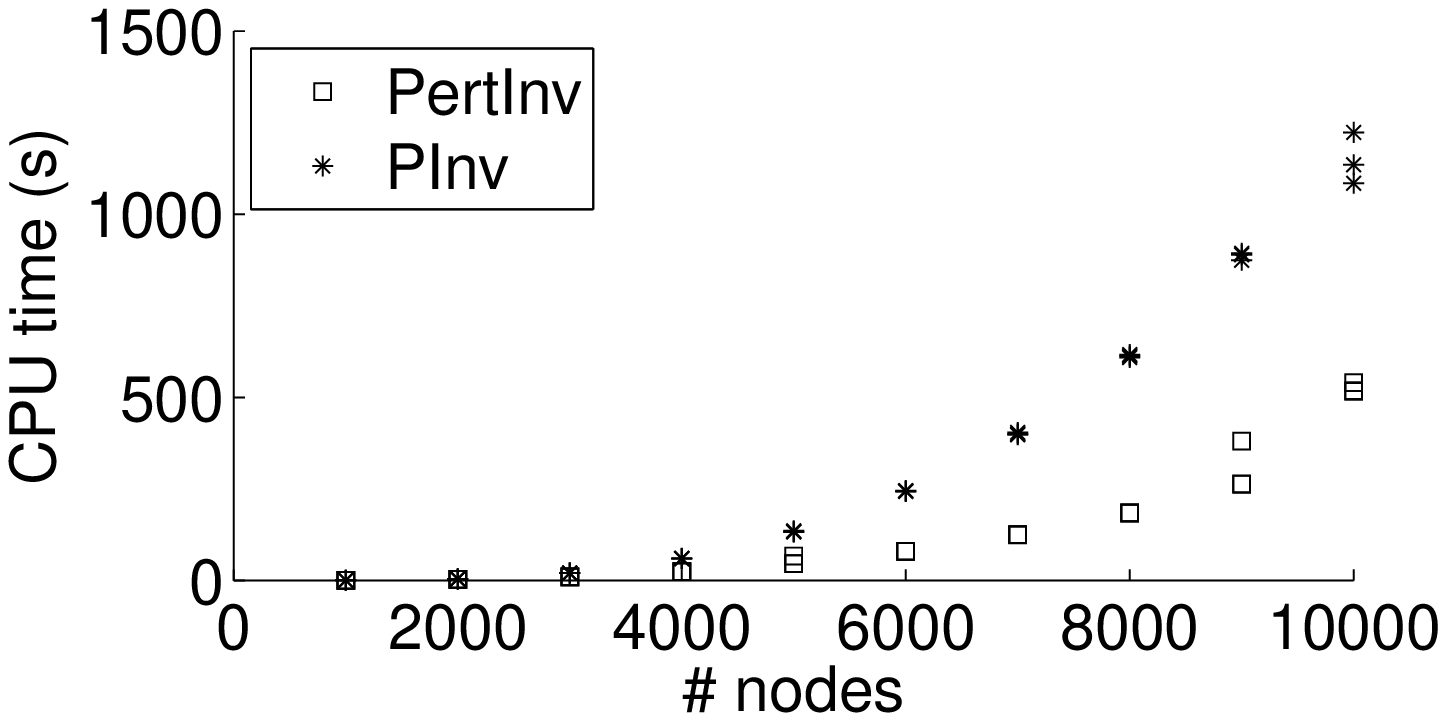}}
&
\scalebox{0.8}{\includegraphics[width = 100mm]{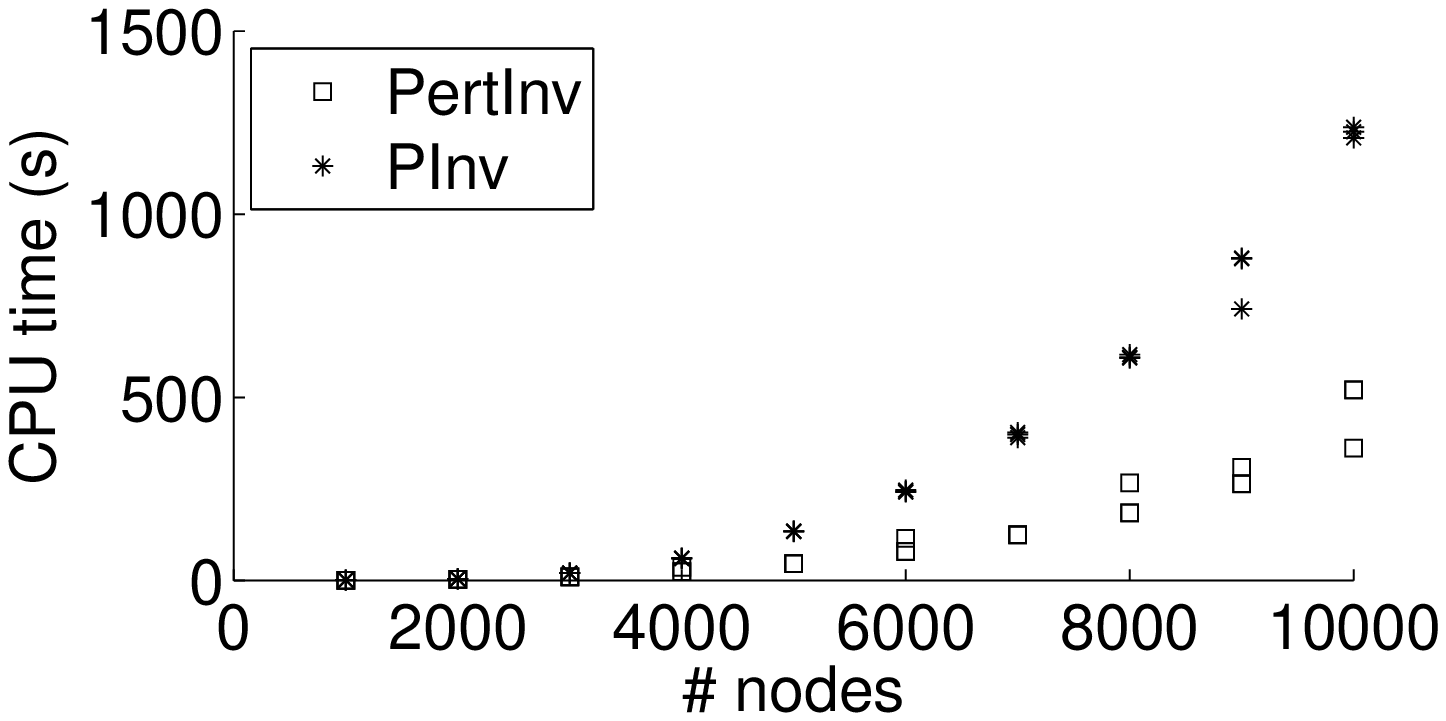}}
\\
(a) $\rho = 0.3$ & (b) $\rho = 0.5$
\end{tabular}}
\caption{Computational times: Erd\"{o}s-R\'{e}nyi graphs of varying orders and densities. 
PertInv: pseudo-inverse computed through $rank(1)$ perturbation \cite{Xiao03}, 
PInv: pseudo-inverse computed through standard $pinv$ in MATLAB.} 
\label{fig:scatterCPUTimeVsOrderForER}
\end{figure}

Fig. \ref{fig:scatterCPUTimeVsOrderForER} shows a comparison of the two methods for 
$\rho = \{0.3, ~0.5\}$ and $n = \{1000, ~2000, ..., ~10000\}$. The experiment is repeated 
100 times for each parametric combination $(n, \rho)$.   
The fact that the method from \cite{Xiao03} outperforms $pinv$ is self evident, as is 
the fact that the computational times for both methods rises with increasing values 
of $n$. We also observe great consistency (or very little variance) across the different 
instances for a given $(n, \rho)$, which is not too surprising.
What is of interest, however, is that the computational times for a given value 
of $n$ do not vary significantly across $\rho = \{0.3, ~0.5\}$, for either of the two methods.   
We observe the same for higher values of $\rho$ (not shown here).  
This implies that the methods are insensitive to the sparsity of the graphs. 
Moreover, for graphs of $(n, ~\rho) = (10000, ~0.5)$, the primary memory imprint for both 
methods is over $2.0 ~GB$ when run in MATLAB 
(a little higher, in fact, for the perturbed inversion method). 
Although the exact figures may vary from machine to machine, they provide a rough estimate 
that suffices for the problem at hand.       
In summary, for dynamically changing graphs, in which small 
local modifications occur every now and then, such methods would incur undue heavy 
computational costs due to repeated re-computation of $\bb L^+$ from scratch.  
On the other hand, for graphs of higher orders ($n > O(10^5)$), 
such decomposition/inversion based methods are rendered impractical from the 
point of view of computational time as well as memory requirements, if 
performed on a single machine.        

In what follows, we show that the computation of the Moore-Penrose pseudo-inverse of the 
Laplacian can be done in a divide-and-conquer fashion. 
Our method allows efficient incremental updates of $\bb L^+$ for dynamically changing graphs, 
without having to compute $\bb L^+$ all over again. Moreover, computing $\bb L^+$ for large 
graphs becomes feasible, in principle, through parallelization of (smaller) independent 
sub-problems over multiple machines, which can then be re-combined at $O(n^2)$ cost per 
edge across a division (details in a subsequent section). 
But first we need to establish a few more preliminary results to further motivate our study.       
\subsection{Sub-Matrix Inverses of $\bb L$}
\label{SubSec:SubInvOfLap}
\ignore{
\begin{figure}[t]
\centerline{\begin{tabular}{c}
\scalebox{0.8}{\includegraphics[width = 120mm]{./dwg_SubMatInv_LpFull.eps}}
\end{tabular}}
\caption{Scalar mapping: Sub-matrix inverse of $\bb L$ to $\bb L^+$.} 
\label{fig:SubMatInvToLpFull}
\end{figure}
}
As described in the previous section, the combinatorial Laplacian $\bb L$ of a connected 
graph $G(V, E)$, is singular and thus non-invertible. However, given that its rank is $n - 1$, 
any $n-1$ combination of columns (or rows)  of $\bb L$ constitutes a linearly independent set.  
Hence, any $(n - 1 \times n - 1)$ sub-matrix of $\bb L$ is invertible.
Indeed, the inverses of such $(n - 1 \times n - 1)$  sub-matrices are made use of in several graph 
analysis problems: enumerating the spanning trees and spanning forests of the graph \cite{Kirkland97}, 
determining the random-walk betweenness of the nodes of the graph \cite{Newman05}, to name a few.  
However, the cost of computing an $(n - 1 ~\times~ n - 1)$ sub-matrix inverse is still $O(n^3)$. 
To compute all such sub-matrix inverses amounts to a time complexity of $O(n^4)$.  
In the following, we show how they can be computed efficiently through $\bb L^+$.     
\begin{mythm}
\label{thm:SubInvByLp}
Let $\bb L(\{\overline n\}, \{\overline n\})$ be an $(n-1 \times n-1)$ sub-matrix of $\bb L$ formed by 
removing the $n^{th}$ row and $n^{th}$ column of $\bb L$. Then $\forall (x, y) \in V(G) \times V(G)$: 
\begin{equation}
[\bb L(\{\overline n\}, \{\overline n\})^{-1}]_{xy} = l^+_{xy} - l^+_{xn} - l^+_{ny} + l^+_{nn}  
\end{equation}
\end{mythm}

\noindent The result in Theorem $\ref{thm:SubInvByLp}$ above expresses, in scalar form, 
the general element ($x^{th}$ row, $y^{th}$ column) of the inverse of the sub-matrix 
$\bb L(\{\overline n\}, \{\overline n\})$ in terms of the elements of $\bb L^+$, as claimed. 
As the choice of the $n^{th}$ row and column is arbitrary, we can see that the result holds 
in general for any $(n-1 \times n-1)$ sub-matrix (permuting the rows and columns of $\bb L$ as 
per need). The cost of computing $\bb L(\{\overline n\}, \{\overline n\})^{-1}$ for a given 
vertex $n$ is $O(n^2)$. Therefore, all sub-matrix inverses can be computed in $O(n^3)$ time from 
$\bb L^+$, which itself can be computed in $O(n^3)$ time, even if the standard methods are used. 
This is clearly an order of magnitude improvement.  Henceforth, we focus entirely on $\bb L^+$.     
\
\subsection{The Effective Resistance Distance and $\bb L^+$}
\label{SubSec:ERDistLp}
\begin{figure}[t]
\centerline{\begin{tabular}{c}
\scalebox{0.80}{\includegraphics[width = 120mm]{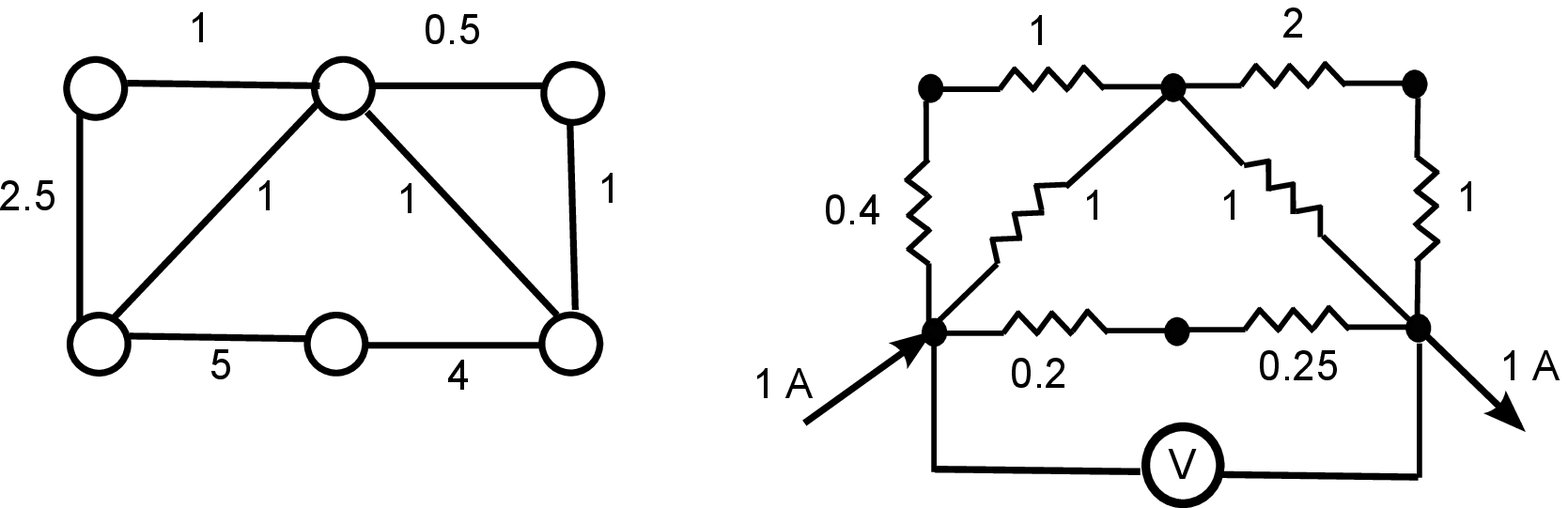}}
\end{tabular}}
\caption{A simple graph $G$  and its EEN.}
\label{fig:GraphEEN}
\end{figure}
An interesting analogy exists between graphs and resistive electrical circuits 
\cite{DoyleSnell84, Fouss07, KleinRandic93}. Given a simple, connected and undirected graph 
$G(V, E)$, the equivalent electrical network $(EEN)$ of the graph can be formed by replacing 
each edge $e_{ij} \in E(G)$, of weight $w_{ij}$ with an electrical resistance 
$\omega_{ij} = w_{ij}^{-1} ~ohm$ (c.f. Fig. \ref{fig:GraphEEN}). 
A distance function can  then be defined between any pair of nodes $(x, y) \in V(G) \times V(G)$ 
in the resulting EEN as follows: 
\begin{mydef}      
Effective Resistance ($\Omega_{xy}$): The voltage developed between nodes x and y, when a unit 
current is injected at node x and is withdrawn at node y.   
\end{mydef}
It is well established that the square root of the effective resistance distance 
($\sqrt{\Omega_{xy}}$) is a Euclidean metric with interesting applications 
\cite{Fouss07, KleinRandic93}.     
Amongst other things, it determines the expected length of random commutes 
between node pairs in the graph: $C_{xy} = vol(G)~ \Omega_{xy}$, ~\cite{Chandra89, Tetali91}. 
More importantly,  $\Omega_{xy}$ can be expressed in terms of the elements 
of $\bb L^+$ as follows:
\begin{equation}
\label{equ:OmegaByLp} 
\Omega_{xy} = l^+_{xx} + l^+_{yy} - l^+_{xy} - l^+_{yx}  
\end{equation}
We now invert the elegant form in ($\ref{equ:OmegaByLp}$) to derive an important result in the 
following lemma which gives us the general term of $\bb L^+$  in terms of the distance function 
$\Omega$.      
\begin{mylem}
\label{lem:SumOmegaTransitTriple}
$\forall (x, y, z) \in V(G) \times V(G) \times V(G):$
\begin{equation} 
l^+_{xy} = \frac{1}{2n}\left(\sum_{z=1}^{n} \Omega_{xz} + \Omega_{zy} - \Omega_{xy}\right) 
- \frac{1}{2n^2} \sum_{x=1}^{n} \sum_{y=1}^{n} \Omega_{xy}  
\end{equation}
\end{mylem}
The $RHS$ in Lemma $\ref{lem:SumOmegaTransitTriple}$ above is composed of two terms: 
a triangle inequality of effective resistances \cite{Tetali91} and a double summand 
over all pairwise effective resistances in the EEN. It is easy to see that the double-summand simply 
reduces to a scalar multiple of the trace of $\bb L^+$ ($Tr(\bb L^+) = \sum_{z = 1}^n l^+_{zz}$). 
Thus the functional half that determines the elements of $\bb L^+$, is the triangle inequality 
of the effective resistances, while the double summand contributes an additive constant 
to all the entries of $\bb L^+$. We illustrate the utility of this result,  with the help of two kinds of 
graphs on the extremal ends of the connectedness spectrum: the star and 
the clique.\footnote{The graphs in these examples are assumed to be unweighted, i.e. all edges 
have a unit resistance/conductance.}   
\subsubsection{The Star}
\label{SubSubSec:Star}
\begin{figure}[t]
\centerline{\begin{tabular}{c}
\scalebox{0.50}{\includegraphics[width = 200mm]{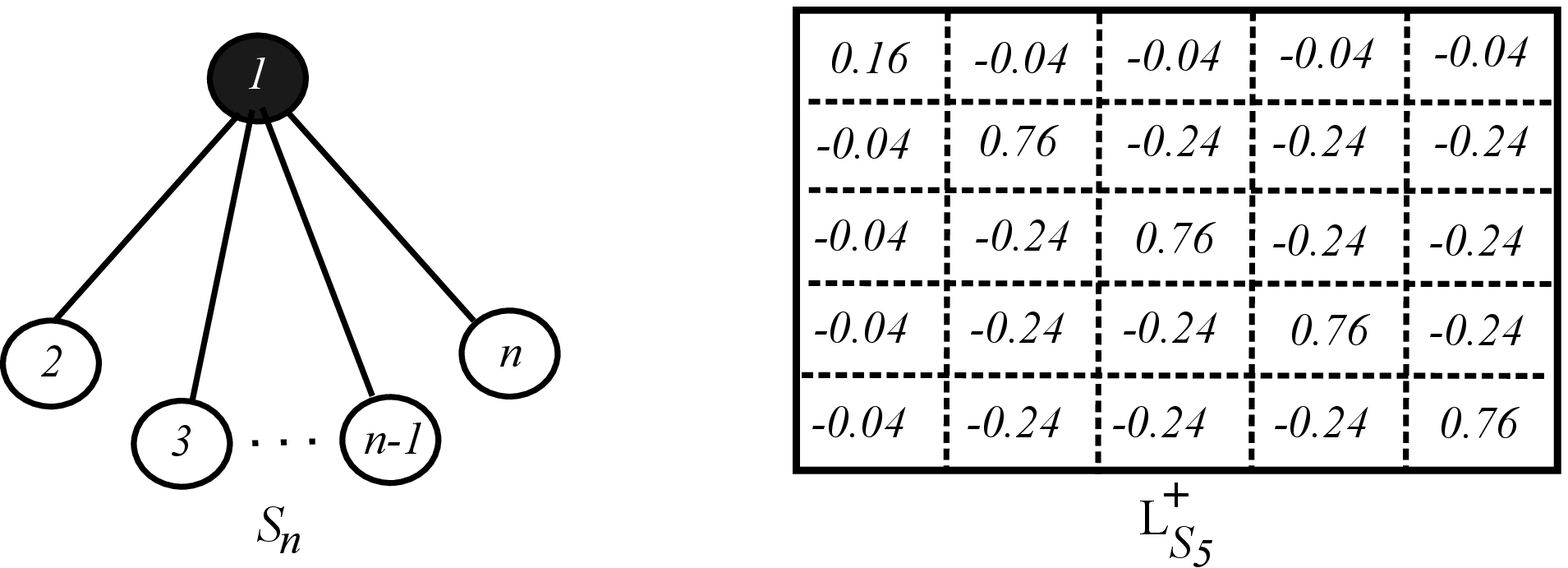}}
\end{tabular}}
\caption{The Star Graph: Pre-computed $\bb L^+_{S_n}$ for $n = 5$.}   
\label{fig:StarLp}
\end{figure}

A star of order $n$ is a tree with exactly one vertex of degree $n - 1$, 
referred to as the {\em root}, and $n - 1$ pendant vertices each of degree $1$, called {\em leaves},    
(c.f. Fig. \ref{fig:StarLp}). By definition, a singleton isolated vertex is also a degenerate star 
albeit with no leaves.  It is easy to see that $S_n$, being a tree, is the most sparse connected graph 
of order $n$ (with exactly $n-1$ edges). Also, $S_n$ is the most compact tree of its order 
(lowest diameter). In the following, we show how $\bb L^+_{S_n}$ can be computed  
using the result of Lemma \ref{lem:SumOmegaTransitTriple}.    
\begin{mycor}
\label{cor:StarLp}
For a star graph $S_n$ of order $n$, with node $v_1$ as root and nodes $\{v_2, v_3, ..., v_n\}$ as leaves, 
$\bb L^+_{S_n}$ is given by:   
\begin{equation}
l^+_{11} = \frac{n-1}{n^2} ~~~~~~~~~~~~~~~~~~~~~~~~~~~~~ and  ~~~~~~~~~~~~~~
\forall x: 2 \leq x \leq n, ~~ 
l^+_{1x} =  l^+_{x1} =  -\frac{1}{n^2}
\end{equation}
\begin{equation}
\forall x: 2 \leq x \leq n, ~~ 
l^+_{xx} = \frac{n^2 - n - 1}{n^2} ~~~~~~ and ~~~~~~~
\forall x \neq y: 2 \leq x, y \leq n,  ~~ 
l^+_{xy} = l^+_{yx} = - \frac{n + 1}{n^2}
\end{equation}
\end{mycor}
\subsubsection{The Clique}
\label{SubSubSec:Clique}
On the other end of the connectedness spectrum lies the clique. A clique $K_n$ of order $n$ is a 
complete graph with $\frac{n (n -1)}{2}$ edges. Clearly, the clique is the densest possible graph 
of order $n$, as there is a direct edge between any pair of vertices in it. It is also the most compact 
graph of its order (lowest diameter).  Then,    
\begin{mycor}
\label{cor:CliqueLp}
For a clique $K_n$ of order $n$, $\bb L^+_{K_n}$ is given by:   
\begin{equation}
\forall x: 1 \leq x \leq n, ~~ l^+_{xx} = \frac{n -1}{n^2} ~~~~~~~~~~~~~~~~~~ and  ~~~~~~~~~~~~~~
\forall x \neq y: 1 \leq x, y \leq n,  ~~ 
l^+_{xy} = l^+_{yx} =  -\frac{1}{n^2}
\end{equation}
\end{mycor}
\noindent The results in the corollaries presented above are not just illustrative examples. They 
are also of interest from a computational point of view, particularly when the graph under study is 
an unweighted one. 
Both stars and cliques can occur as motif sub-graphs in any given graph. 
Indeed, for any non-trivial connected simple graph of order 
$n \geq 3$, there is at least one sub-graph that is a star. Selecting any vertex $i$ with $d(i) \geq 2$, 
and conducting a one-hop breadth first search, generates a star sub-graph. Cliques, though not 
so universal, also occur in real world networks (e.g. citation networks). Therefore, in any 
divide-and-conquer methodology, both stars and cliques are likely to be found at some stage. 
We have already established that the cost of computing 
$\bb L^+_{S_n}$ and $\bb L^+_{K_n}$ is $O(1)$ (as they are determined entirely by $n$) 
and hence the solution to such a sub-problem, when found, is obtained at the lowest possible 
cost --- a true practical gain.     

To conclude, we have demonstrated that there exists a relationship between the elements of 
$\bb L^+$ and the pairwise effective resistances in the graph $G(V, E)$, that yields interesting 
closed form solutions for the pseudo-inverse for special graphs such as stars and cliques.  
In the subsequent sections, we demonstrate that it can be used to compute $\bb L^+$ for general 
graphs as well, incrementally, in a divide-and-conquer fashion. 
\
\

%
%

\section{From Two to One: Computing $\bb L^+$ by Partitions} 
\label{Sec:TwoToOne}
\begin{figure}[t]
\centerline{\begin{tabular}{c}
\scalebox{0.75}{\includegraphics[width = 200mm]{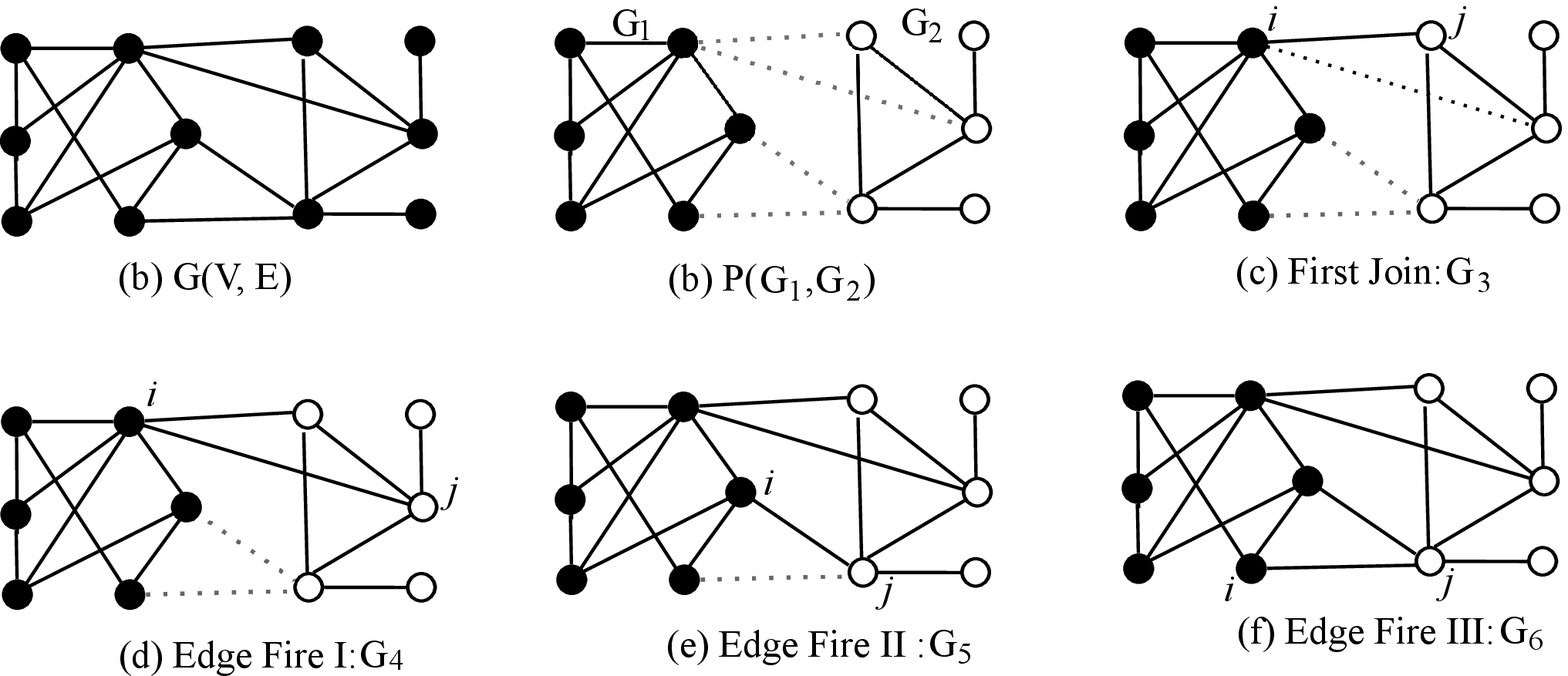}}
\end{tabular}}
\caption{Divide-and-Conquer: Connected bi-partition of a graph and the two-stage process: 
first join followed by three edge firings. The dotted lines represent the edges that are not part 
of the intermediate graph at that stage.}  
\label{fig:TwoStageTwoToOne}
\end{figure}
In this section, we present our main result --- the computation of the Moore-Penrose
pseudo-inverse of the Laplacian, or $\bb L^+$, by means of graph bi-partitions.  
In \textsection\ref{SubSec:TwoStage}, we lay out a {\em two-stage} process 
 --- the {\em first join} followed by {\em edge firings} ---  that underpins 
our methodology. We then provide specific closed form solutions  
in \textsection\ref{SubSec:TwoStageAndLp}. 
\subsection{Connected Bi-Partitions of a Graph and the Two-Stage Process}
\label{SubSec:TwoStage}
In order to compute the Moore-Penrose pseudo-inverse of the Laplacian of a simple, 
connected, undirected and unweighted graph $G(V, E)$ by parts,  
we must first establish that the problem can be decomposed into two, 
or more, sub-problems that can be solved independently. 
The solutions to the independent sub-problems can then be combined to obtain the 
overall result.  
But before we proceed to do so, a few notations are in order.  
\begin{mydef}
Connected Bi-partition $(P=(G_1, G_2))$: A cut of the graph $G$ which contains exactly two 
mutually exclusive and exhaustive connected sub-graphs $G_1$ and $G_2$.
\end{mydef}
Fig. \ref{fig:TwoStageTwoToOne}(a-b), shows a graph $G(V, E)$ and a 
connected bi-partition $P(G_1, G_2)$ of it, obtained from the graph $G(V, E)$ 
by removing the set of dotted edges shown. Each partition $P(G_1, G_2)$ 
has certain defining characteristics in terms of the set of vertices 
as well the set of edges in the graph.   
Let, $V_1(G_1)$ and $V_2(G_2)$ be the mutually exclusive and exhaustive subsets of 
$V(G)$ i.e. $V_1(G_1) \cap V_2(G_2) = \phi$ and $V_1(G_1) \cup V_2(G_2) = V(G)$. 
Similarly, let $E_1(G_1)$ and $E_2(G_2)$ be the sets of edges in the respective sub-graphs 
$G_1$ and $G_2$ of $P$ and $E(G_1, G_2)$, the set of edges that {\em violate} the 
partition $P(G_1, G_2)$ i.e. have one end in $G_1$ and the other in $G_2$. 
Thus, $E_1(G_1) \cap E_2(G_2) = E_1(G_1) \cap E_1(G_1, G_2) = E(G_1, G_2) \cap E_2(G_2) = \phi$
and $E_1(G_1) \cup E(G_1, G_2) \cup E_2(G_2) = E(G)$.   
We denote by $\mathcal{P}(G)$, the set of all such connected bi-partitions of 
the graph $G(V, E)$.  

It is easy to see that for an arbitrary connected bi-partition $P(G_1, G_2) \in \mathcal{P}(G)$ 
both $G_1$ and $G_2$ are themselves simple, connected, undirected and unweighted  
graphs. Hence, the discussion in \textsection\ref{Sec:LapSubInvDist} is 
applicable in its entirety to the sub-graphs $G_1$ and $G_2$ independently.   
Note then that $\bb L^+_{G_1}$ and $\bb L^+_{G_2}$, 
the Moore-Penrose pseudo-inverse of the Laplacians of the sub-graphs $G_1$ and $G_2$, 
must, by definition, exist. The pair $\{\bb L^+_{G_1}$, $\bb L^+_{G_2}\}$, constitutes 
the solution to two independent sub-problems represented in the set $\{G_1, G_2\}$.  
All that remains to be shown now is that $\{\bb L^+_{G_1}$, $\bb L^+_{G_2}\}$ can indeed be 
combined to obtain $\bb L^+_G$. 
It is this aspect of the methodology, that we call the {\em two-stage} process, as explained in 
detail below.        

The original graph $G(V, E)$ can be thought of, in some sense, as a bringing 
together of the disjoint spanning sub-graphs $G_1$ and $G_2$, by means of 
introducing the edges of the set $E(G_1, G_2)$.    
Starting from $G_1$ and $G_2$, we iterate over the set of edges 
in $E(G_1, G_2)$ in the following fashion 
(c.f. Fig. \ref{fig:TwoStageTwoToOne} for a visual reference).    
Let $e_{ij} \in E(G_1, G_2): i \in V_1(G_1), j \in V_2(G_2)$, of weight $w_{ij}$ 
and resistance $\omega_{ij} = w_{ij}^{-1} ~ohm$, be an arbitrary edge 
chosen during the first iteration as shown in Fig. \ref{fig:TwoStageTwoToOne}(c). 
We call this step the {\em first join} in our two-stage process, whereafter $G_1$ 
and $G_2$ come together to give an intermediate connected spanning sub-graph 
(say $G_3(V_3, E_3)$). 
The first join represents a point of singularity in the reconstruction process, particularly 
from the perspective of the effective resistance distance. 
Note that before the first join, the effective resistance distance between an 
arbitrary pair of nodes $(x, y) \in V(G) \times V(G)$ is infinity, 
if $x \in V_1(G_1)$ and $y \in V_2(G_2)$, as there is no path connecting $x$ and $y$. 
However,  once the first edge $e_{ij}$ has been introduced during the first join, 
this discrepancy no longer exists and all pairwise effective resistances are finite.  
Precisely, if $\Omega^{G_1}: V_1(G_1) \times V_1(G_1) \rightarrow \Re^+$   
and  $\Omega^{G_2}: V_2(G_2) \times V_2(G_2) \rightarrow \Re^+$,  
be the pairwise effective resistances defined over the sub-graphs 
$G_1$ and $G_2$, the following holds:
\begin{eqnarray*}
\label{equArr:FirstJoin}
\Omega^{G_3}_{xy} 
&=& \Omega^{G_1}_{xy}, ~~~~~~ ~~~~~~~ ~~~~~~~~ ~~if ~x, y \in V_1(G_1) \\
&=& \Omega^{G_2}_{xy}, ~~~~~~ ~~~~~~~ ~~~~~~~~ ~~if ~x, y \in V_2(G_2) \\
&=&  \Omega^{G_1}_{xi} + \omega_{ij} +  \Omega^{G_2}_{jy}, ~~~~~~ if ~x \in V_1(G_1) ~~\&~~ y \in V_2(G_2) 
\end{eqnarray*}
Needless to say, this is a critical step in the process as we need finite values of effective 
resistances in order to exploit the result in Lemma $\ref{lem:SumOmegaTransitTriple}$. 
Hereafter, we can combine the solutions to the independent sub-problems, i.e. 
$\bb L^+_{G_1}$ and $\bb L^+_{G_2}$, to obtain $\bb L^+_{G_3}$.
Indeed, we obtain an elegant scalar form with interesting properties (details in subsequent sections). 

Following the first join, the remaining edges in $E(G_1, G_2)$, can now be introduced 
one at a time to obtain a sequence of intermediate graphs 
$\left(G_4 \rightarrow G_5 \rightarrow G_6\right)$  which finally ends in $G(V, E)$ 
(c.f. Fig. \ref{fig:TwoStageTwoToOne}(d-f)). 
We call this second stage of edge introductions, following the first join, {\em edge firing}.  
In terms of effective resistances, each edge firing simply creates parallel resistive connections, 
or alternative paths, in the graph.   
Algebraically, each edge firing is a $rank(1)$ perturbation of the Laplacian for the 
intermediate graph from the previous step. 
Thus, the Moore-Penrose pseudo-inverse of the Laplacians for the 
intermediate graph sequence $\left(G_4 \rightarrow G_5 \rightarrow G_6\right)$ can be obtained 
starting from $\bb L^+_{G_3}$ using standard perturbation methods \cite{Meyer73} 
(details in subsequent sections).              

To summarize, therefore, during the two-stage process we obtain a sequence of connected 
spanning sub-graphs of $G(V, E)$ starting from a partition $P(G_1, G_2) \in \mathcal{P}(G)$, 
performing the first join by arbitrarily selecting an edge $e_{ij} \in E(G_1, G_2)$, and then firing the 
remaining edges, one after the other, in any arbitrary order. 
The number of connected spanning sub-graphs of $G(V, E)$ constructed during the two-stage 
process is exactly $|E(G_1, G_2)|$ ($= 4$ for the example in Fig. \ref{fig:TwoStageTwoToOne}).  
Note that, the order in which these sub-graphs are generated, 
is of no consequence whatsoever.  
Next, we use these insights to obtain $\bb L^+$ for the intermediate graphs in the sequence.    
\subsection{The Two-Stage Process and $\bb L^+$}
\label{SubSec:TwoStageAndLp}
We now present the closed form solutions for the Moore-Penrose pseudo-inverse of the Laplacians 
of the set of intermediate graphs obtained during the two-stage process. 
\subsubsection{The First Join}
\label{SubSubSec:FirstJoin}
\begin{figure}[t]
\centerline{\begin{tabular}{c}
\scalebox{0.60}{\includegraphics[width = 200mm]{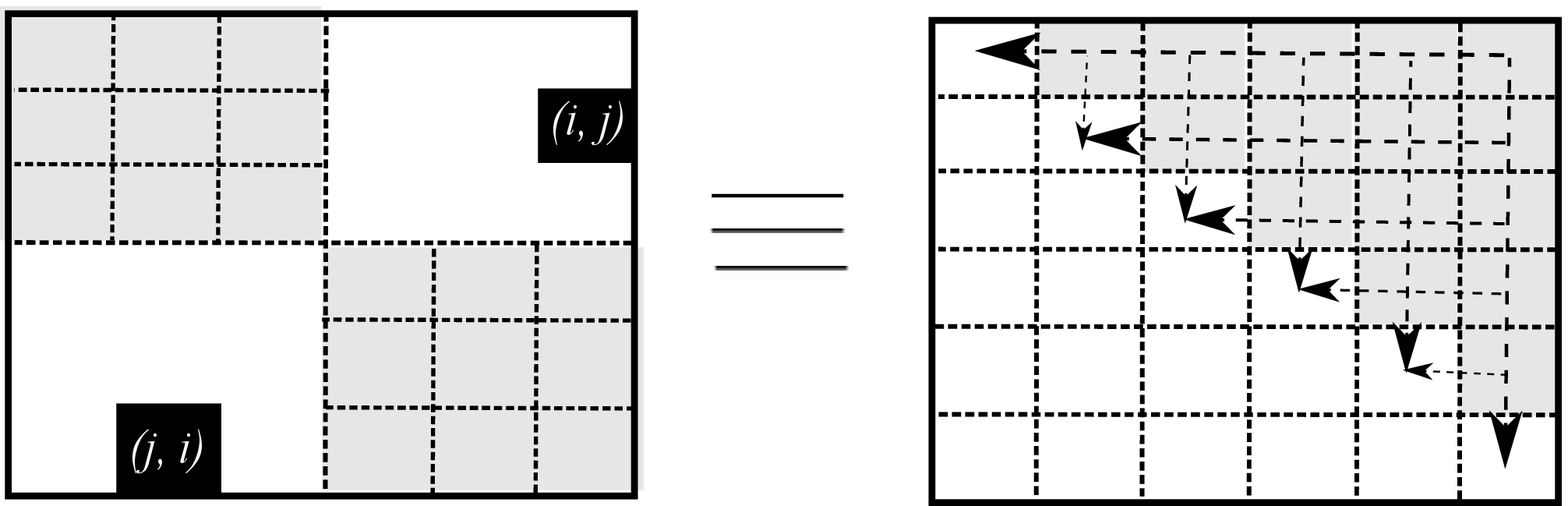}}
\end{tabular}}
\caption{The First Join: Scalar mapping $\left(\bb L^+_{G_1}, ~ \bb L^+_{G_2}\right)$ to $\bb L^+_{G_3}$. 
The grey blocs represent relevant elements in $\bb L^+_{G_1}, ~ \bb L^+_{G_2}$ and $\bb L^+_{G_3}$.  
Arrows span the elements of the upper triangular of $\bb L^+_{G_3}$ that contribute to the respective 
diagonal element pointed to by the arrow head: 
$\displaystyle l^{+(3)}_{kk} = - \left(\sum_{i = 1}^{k-1} l^{+(3)}_{ik} + \sum_{j = k + 1}^n l^{+(3)}_{kj}\right)$.}
\label{fig:FirstJoinLpMatInd}
\end{figure}
Given, two simple, connected, undirected graphs $G_1(V_1, E_1)$ and $G_2(V_2, E_2)$ 
let $\bb L^+_{G_1}$  and $\bb L^+_{G_2}$, be the respective Moore-Penrose pseudo-inverses of 
their Laplacians. Also, let $n_1 = |V_1(G_1)|$ and $n_2 = |V_2(G_2)|$ be the orders of the two graphs. 
We denote by $l^{+(1)}_{xy}$ and $l^{+(2)}_{xy}$ respectively the general terms of the matrices  
$\bb L^+_{G_1}$ and $\bb L^+_{G_2}$. 
Next, let the {\em first join} between $G_1$ and $G_2$ be performed by introducing an edge 
$e_{ij}$ between the graphs $G_1$ and $G_2$ to obtain $G_3(V_3, E_3)$; 
where $i \in V_1(G_1)$ and $j \in V_2(G_2)$. 
Clearly,  $V_3(G_3) = V_1(G_1) \cup V_2(G_2)$ and  
$E_3(G_3) = E_1(G_1) \cup \{e_{ij}\} \cup E_2(G_2)$. 
Thus, $|V_3 (G_3)| = n_3 = n_1 + n_2$ and $E_3(G_3) = m_3 = m_1 + 1 + m_2$. 
By convention, the vertices in $V_3(G_3)$ are labeled in the following order:  
the first $n_1$ vertices $\{1, 2, ..., n_1\}$ are retained, {\em as is}, from $V_1(G_1)$ 
and the remaining $n_2$ vertices are labelled $\{n_1 + 1, n_1 + 2, ..., n_1 + n_2\}$ 
in order from $V_2(G_2)$.   
We denote by $\bb L^+_{G_3}$ the pseudo-inverse and $l^{+(3)}_{xy}$ its general term.   
Then, 
\begin{mythm}
\label{thm:FirstJoinLp}
$\forall (x, y) \in V_3(G_3) \times V_3(G_3)$, 
\begin{eqnarray*}
l^{+(3)}_{xy}
&=& l^{+(1)}_{xy} - 
\frac{n_2 n_3 \left(l^{+(1)}_{xi} + l^{+(1)}_{iy}\right) - n_2^2 \left(l^{+(1)}_{ii} + l^{+(2)}_{jj} + \omega_{ij}\right)}
{n_3^2}, ~~~~~~~~~~ if ~x, y \in V_1(G_1) \\
&=& l^{+(2)}_{xy} - 
\frac{n_1n_3 \left(l^{+(2)}_{xj} + l^{+(2)}_{jy}\right) - n_1^2 \left(l^{+(1)}_{ii} + l^{+(2)}_{jj} + \omega_{ij}\right)}
{n_3^2}, ~~~~~~~~~~ if ~x, y \in V_2(G_2) \\
&=& ~~~~~~~~~~\frac{n_3 \left( n_1l^{+(1)}_{xi} + n_2 l^{+(2)}_{jy}\right) - n_1 n_2 \left(l^{+(1)}_{ii} + l^{+(2)}_{jj} + \omega_{ij}\right)} 
{n_3^2}, ~~~ if ~x \in V_1(G_1) ~~\&~~ y \in V_2(G_2) 
\end{eqnarray*}
\end{mythm}   
The result in Theorem $\ref{thm:FirstJoinLp}$ is interesting for several reasons.  
First and foremost, it clearly shows that the general term of $\bb L^+_{G_3}$,  
is a linear combination of the elements of $\bb L^+_{G_1}$ and $\bb L^+_{G_2}$. 
This was indeed our principal claim. 
Secondly, $\forall (x, y) \in V_3(G_3) \times V_3(G_3)$, each individual $l^{+(3)}_{xy}$ can 
be computed independent of the others 
(barring symmetry, i.e. $l^{+(3)}_{xy} = l^{+(3)}_{yx}$, which we shall discuss shortly).  
They are determined entirely by the specific elements from the $i^{th}$ and $j^{th}$ columns 
of the matrices  $\bb L^+_{G_1}$ and $\bb L^+_{G_2}$, depending upon the membership of  
$x$ and $y$ in the disjoint graphs. This implies that all $l^{+(3)}_{xy}$ can be computed in parallel, 
as long as we have the relevant elements of $\bb L^+_{G_1}$ and $\bb L^+_{G_2}$.   

From a cost point of view, the first join requires $O(1)$ computations  
per element in $\bb L^+_{G_3}$ --- 
constant number of $\{+, ~-, ~\times , ~/\}$ operations --- 
if $\{\bb L^+_{G_1}, ~\bb L^+_{G_2}\}$ is given {\em a priori}.   
The common term in the numerator, i.e. $(l^{+(1)}_{ii} + l^{+(2)}_{jj} + 1)$, is an invariant 
for the elements of $\bb L^+_{G_3}$ and need only be computed once.    
This term is simply a linear multiple of the change in trace:   
\begin{equation}
\Delta(Tr) =  Tr(\bb L^+_{G_3}) - \left[Tr(\bb L^+_{G_1}) + Tr(\bb L^+_{G_2})\right]
\end{equation}
For details see the proof of Lemma \ref{lem:TraceByParts} in Appendix. Therefore, 
we achieve an overall cost of $O(n_3^2)$ for the first join.   
Last but not the least, we need to compute and store only the upper triangular of 
$\bb L^+_{G_3}$. Owing to the symmetry of $\bb L^+_{G_3}$, the lower triangular  
is determined automatically. As for the diagonal elements, they come without any additional 
cost as a result of $\bb L^+_{G_3}$ being doubly-centered (c.f. Fig. \ref{fig:FirstJoinLpMatInd}).  
\subsubsection{Firing an Edge}
\label{SubSubSec:FiringEdge}
We now look at the second stage that of {\em firing an edge} in a connected graph.    
Given a simple, connected, undirected graph $G_1(V_1, E_1)$, let  $e_{ij} \notin E_1(G_1)$  
be {\em fired} to obtain $G_2(V_2, E_2)$. 
Clearly, $V_2(G_2) = V_1(G_1)$ and $E_2(G_2) = E_1(G_1) \cup \{e_{ij}\}$. 
Continuing with our convention, we denote by $\bb L^{+}_{G_1}$ and  $\bb L^{+}_{G_2}$ the 
Moore-Penrose pseudo-inverses of the respective Laplacians. Then, 
\begin{mythm}  
\label{thm:EdgeFireLp}
$\forall (x, y) \in V_2(G_2) \times V_2(G_2)$,  
\begin{equation}
l^{+(2)}_{xy} 
= l^{+(1)}_{xy} 
- \frac{\left(l^{+(1)}_{xi} - l^{+(1)}_{xj}\right) \left(l^{+(1)}_{iy} - l^{+(1)}_{jy}\right)}{\omega_{ij} + \Omega^{G_1}_{ij}} 
\end{equation}
\end{mythm}
where $\Omega^{G_1}_{ij}$ is the effective resistance distance between nodes $i$ and $j$ in the graph 
$G_1(V_1, E_1)$ --- an invariant $\forall (x, y) \in V_3(G_3) \times V_3(G_3)$ that is determined entirely 
by the end-points of the edge $e_{ij}$ being fired. 
Once again, we observe that the general term of $\bb L^+_{G_2}$ is a 
linear combination of the elements of $\bb L^+_{G_1}$ and requires $O(1)$
 computations per element in $\bb L^+_{G_2}$ --- 
constant number of $\{+, ~-, ~\times , ~/\}$ operations --- 
if $\bb L^+_{G_1}$ is given {\em a priori}. 
The rest of the discussion from the preceding sub-section on first join ---  
element-wise independence and  upper triangular sufficiency --- holds {\em as is} 
for this stage too.  However, before concluding this section, we extend the result of 
Theorem \ref{thm:EdgeFireLp} to the pairwise effective resistances themselves  
in the following corollary.  
\begin{mycor}  
\label{cor:EdgeFireOmega}
$\forall (x, y) \in V_2(G_2) \times V_2(G_2)$,  
\begin{equation}
\Omega^{G_2}_{xy} 
= \Omega^{G_1}_{xy} 
- \frac{\left[\left(\Omega^{G_1}_{xj} - \Omega^{G_1}_{xi}\right) - \left(\Omega^{G_1}_{jy} - \Omega^{G_1}_{iy}\right)\right]^2}
{4(\omega_{ij} + \Omega^{G_1}_{ij})} 
\end{equation}
\end{mycor}
The result above is interesting in its own right. Note that computing $\Omega^{G_2}$ 
when the edge density of a graph increases 
(or the expected commute times in random walks between nodes),   
is pertinent to many application scenarios 
\cite{Brand05, Chandra89, FoussPirotteRendersSaerens05, FoussYenPirotteSaerens06, Fouss07, 
Luxburg10, SaerensFoussYenDupont04, SarwarEtAl02}.  
Corollary \ref{cor:EdgeFireOmega} gives us a way of computing these distances directly 
without having to compute $\bb L^+_{G_2}$ first.   

To conclude, therefore, we have established that the Moore-Penrose pseudo-inverses of the 
Laplacians of all the intermediate graphs, generated during the two-stage process, are incrementally 
computable from the solutions at the preceding stage, on an element-to-element basis. 
We shall return to specific applications of these results to dynamic (time-evolving) graphs and 
large graphs in general, in a subsequent section. 
But first, for the sake of completeness, we present the case of structural regress.      
\

%
%

%
%

\section{From One to Two: A Case of Regress}
\label{Sec:OneToTwo}
We now present analogous results in the opposite direction, that of structural regress of a graph 
through successive deletion of edges until the graph breaks into two. These results, similar in essence to those 
presented in the preceding section, are particularly significant with respect to 
dynamically evolving graphs that change with time (e.g. social networks). Once again, 
we have two cases to address with respect to edge deletions {\em viz.}  
(a) {\em Non-bridge edge}: an edge that upon deletion does not affect the connectedness of the graph   
(c.f. \textsection\ref{SubSec:DelNonBridgeEdge}); and 
(b) {\em Bridge-edge}: an edge that, when deleted, yields a connected bi-partition of the graph 
(c.f. \textsection\ref{SubSec:DelBridgeEdge}).  
\subsection{Deleting a Non-Bridge Edge}
\label{SubSec:DelNonBridgeEdge}
Given a simple, connected, undirected graph $G_1(V_1, E_1)$, let $e_{ij} \in E_1(G_1)$  
be a {\em non-bridge} edge that is deleted to obtain $G_2(V_2, E_2)$. 
Clearly, $V_2(G_2) = V_1(G_1)$ and $E_2(G_2) = E_1(G_1) - \{e_{ij}\}$. 
Once again, we denote by $\bb L^{+}_{G_1}$ and  $\bb L^{+}_{G_2}$ the 
Moore-Penrose pseudo-inverses of the respective Laplacians. Then, 
\begin{mythm}
\label{thm:NonBridgeEdgeDelLp}
$\forall (x, y) \in V_2(G_2) \times V_2(G_2)$,
\begin{equation}
l^{+(2)}_{xy} = 
l^{+(1)}_{xy} + \frac{\left(l^{+(1)}_{xi} - l^{+(1)}_{xj}\right) \left(l^{+(1)}_{iy} - l^{+(1)}_{jy}\right)}{\omega_{ij} - \Omega^{G_1}_{ij}} 
\end{equation}
\end{mythm} 
Note, as $e_{ij}$ is a non-bridge edge, $\Omega^{G_1}_{ij} \neq 1$. 
In fact, given that $G_1(V_1, E_1)$ is connected, undirected and unweighted, we have: 
$0 < \Omega^{G_1}_{ij} < 1$. 
Also, as in the case of the two-stage process, we observe the same element-wise independence for 
$\bb L^+_{G_2}$ here as well. Once again, if the quantity of interest is $\Omega^{G_2}$ or 
pairwise expected commute times in random walks, we can simply use the following corollary. 
\begin{mycor}  
\label{cor:NonBridgeEdgeDelOmega}
$\forall (x, y) \in V_2(G_2) \times V_2(G_2)$,  
\begin{equation}
\Omega^{G_2}_{xy} 
= \Omega^{G_1}_{xy} 
+ \frac{\left[\left(\Omega^{G_1}_{xj} - \Omega^{G_1}_{xi}\right) 
- \left(\Omega^{G_1}_{jy} - \Omega^{G_1}_{iy}\right)\right]^2}
{4(\omega_{ij} - \Omega^{G_1}_{ij})} 
\end{equation}
\end{mycor}
\subsection{Deleting a Bridge Edge}
\label{SubSec:DelBridgeEdge}
Finally, we deal with the case when a bridge edge is 
deleted from a graph, thus rendering it disconnected for the first time. 
This represents the point of singularity in the case of structural regress 
(analogous to the first join).   
Continuing with our convention, let $G_1(V_1, E_1)$ be a simple, 
connected, undirected graph with a bridge edge $e_{ij} \in E_1(G_1)$. 
Upon deleting $e_{ij}$, we obtain $G_2(V_2, E_2)$ and $G_3(V_3, E_3)$, 
two disjoint spanning sub-graphs of $G_1$. The orders of $G_1$, $G_2$ and 
$G_3$ are respectively given by $n_1$, $n_2$ and $n_3$, while  
$\bb L^+_{G_1}, \bb L^+_{G_2}$ and $\bb L^+_{G_3}$ are the respective 
pseudo-inverse matrices of their Laplacians. 
It is easy to see that: 
\begin{eqnarray*}
\label{equArr:BridgeDelSameSide}
\Omega^{G_1}_{xy} 
&=& \Omega^{G_2}_{xy}, ~~~~~~ ~~~~~~ ~~~~~~ ~~if ~x, y \in V_2(G_2) \\
&=& \Omega^{G_3}_{xy}, ~~~~~~ ~~~~~~ ~~~~~~ ~~if ~x, y \in V_3(G_3) \\
\end{eqnarray*}
and $\Omega^{G_2 \times G_3}_{xy} =  \Omega^{G_3 \times G_2}_{xy} = \infty$, as $G_1$ and 
$G_2$ are disjoint. 
To obtain $\bb L^+_{G_2}$ and $\bb L^+_{G_3}$ from $\bb L^+_{G_1}$, we 
use the result in Lemma $\ref{lem:SumOmegaTransitTriple}$. 
\begin{mythm}
\label{thm:BridgeEdgeDelLp}
$\forall (x, y) \in V_2(G_2) \times V_2(G_2)$ and $\forall (u, v) \in V_3(G_3) \times V_3(G_3)$,
\begin{equation}
\label{equ:BridgeEdgeDelLpG2}
l^{+(2)}_{xy} = 
l^{+(1)}_{xy} - 
\frac{
\displaystyle n_2 \sum_{z \in V_2(G_2)} 
\left(l^{+(1)}_{xz} + l^{+(1)}_{zy}\right)
- \sum_{x \in V_2(G_2)}\sum_{y \in V_2(G_2)} l^{+(1)}_{xy}}
{n_2^2} 
\end{equation}
\begin{equation}
\label{equ:BridgeEdgeDelLpG3}
l^{+(3)}_{uv} = 
l^{+(1)}_{uv} - 
\frac{
\displaystyle n_3 \sum_{w \in V_3(G_3)} 
\left(l^{+(1)}_{uw} + l^{+(1)}_{wv}\right)
- \sum_{u \in V_3(G_3)}\sum_{v \in V_3(G_3)} l^{+(1)}_{uv}}
{n_3^2} 
\end{equation}
\end{mythm} 
Note also that $\bb L^+_{G_2} \in \Re^{n_2 \times n_2}$  and  
$\bb L^+_{G_2} \in \Re^{n_3 \times n_3}$. For convenience, 
and without loss of generality, we assume that the rows and columns of 
$\bb L^+_{G_1} \in \Re^{n_1 \times n_1}$ have been pre-arranged in such a way that the 
first $(n_2 \times n_2)$ sub-matrix (upper-left) maps to the sub-graph $G_2$ and 
similarly the lower-right $(n_3 \times n_3)$ sub-matrix to $G_3$. 
\

%
%

\section{Bringing it together: Algorithm, Complexity and Parallelization}
\label{Sec:CompAndParallel}
In this section, we bring together the results obtained in \textsection\ref{Sec:TwoToOne} 
and \textsection\ref{Sec:OneToTwo}, to bear on two important scenarios:  
(a) dynamic ( time-evolving) graphs (c.f. \textsection\ref{SubSec:DynGraph}), and 
(b) real-world networks of large orders   
(c.f. \textsection\ref{SubSec:DivAndConq}). In each case, we discuss the time complexity 
and parallelizability of our approach in detail. 
\subsection{Dynamic Graphs: Incremental Computation for Incremental Change}
\label{SubSec:DynGraph}
Dynamic graphs are often used to represent temporally changing systems. The most intuitively 
accessible example of such a system is an online social network (OSN). 
An OSN evolve not only in terms of order, through introduction and attrition of users with time,  
but also in terms of the social ties (or relationships) between the users as new associations are formed, 
and older ones may fade off. Mathematically, we model an OSN as a dynamic graph 
$G_\tau(V_\tau, E_\tau)$ where the sub-index $\tau$ 
denotes the time parameter. We now study a widely used model for dynamic, temporally 
evolving, graphs called {\em preferential attachment} \cite{Barabasi00, Barabasi99, Barabasi01}.      

The preferential attachment model is a parametric model for network growth determined 
by parameters $(n, \kappa)$ such that $n$ is the desired order of the network and $\kappa$ 
is the desired average degree per node. In its simplest form, the model 
proceeds in discrete time steps whereby at each time instant $1 < \tau+1 \leq n$, 
a new node $v_{\tau+1}$ is introduced in the network with $\kappa$ edges. This incoming node 
$v_{\tau+1}$, gets attached to a node $v_i: 1 \leq i \leq \tau$, through exactly one of its $\kappa$ edges, 
with the following probability: 
\begin{equation}
\label{equ:PAProbAttach}
P_{\tau + 1}(v_i) = \displaystyle \frac{d_\tau(i)}{\sum_{j = 1}^{\tau} d_\tau(j)}
\end{equation} 
where $d_\tau(i)$ is the degree of node $v_i$ at time $\tau$. The end-points of all the edges 
emanating from $v_{\tau + 1}$ are selected in a similar fashion. 
At the end of time step $\tau + 1$, we obtain $G_{\tau+1}(V_{\tau + 1}, E_{\tau + 1})$, and the process 
continues until we have a graph $G_n(V_n, E_n)$ of order $n$.\footnote{In practice, for $\kappa > 1$, 
the process starts with a small connected network as a base substrate to facilitate probabilistic selection 
of neighbors for an incoming node. For $\kappa = 1$, we may start with a singleton node, and the resulting 
structure is a tree.}          

Simplistic though it may seem, this model has been shown to account for several characteristics observed 
in real-world networks, including the {\em power law} degree distributions, the {\em small-world} characteristics 
and the logarithmic growth of network diameter with time \cite{Barabasi00, Barabasi99, Barabasi01}. 
We return to these in detail in the next sub-section while dealing with the more general case. 

It is easy to see that in order to study the structural evolution of dynamic networks,  
particularly in terms of the sub-structures like spanning trees and forests \cite{Kirkland97}, 
or centralities of nodes and edges \cite{Newman05, RanjanZhang11a}; or voltage distributions 
in growing conducting networks \cite{Tadic02}, we require not only the final state $G_n(V_n, E_n)$, 
but all the intermediate states of the network.  
In other words, we need to compute the pseudo-inverses of the Laplacians for all the graphs in the 
sequence $(G_{1} \rightarrow G_{2} \rightarrow ... \rightarrow G_{n})$. 
Clearly, if the standard methods are used, the cost at time step $\tau$ is $O(\tau^3)$. 
The overall asymptotic cost for the entire sequence is then 
$O\left(\displaystyle \sum_{\tau = 1}^{n} \tau^3 = \left[\frac{n (n+1)}{2}\right]^2\right)$. 

In contrast, using our incremental approach, we can accomplish this at a much lower computational cost. 
Note that in the case of growing networks, we do not need an explicit divide operation 
at all. The two sub-problems at time step  $\tau + 1$  are given {\em a priori}. 
We have, $G_{\tau}(V_\tau, E_\tau)$  and a singleton vertex graph $\{v_{\tau + 1}\}$ as a pair of 
disjoint sub-graphs. The $\kappa$ edges emanating from $\{v_{\tau + 1}\}$  
have end-points in $G_{\tau}$ as determined by ($\ref{equ:PAProbAttach}$).   
The conquer operation is then performed through a first join between the singleton 
node $\{v_{\tau + 1}\}$ and the graph $G_{\tau}(V_\tau, E_\tau)$. 
We can assume that $\bb L^+_{G_\tau}$ is already known at this time step (the induction hypothesis). 
Also, $\bb L^+_{\{v_{\tau + 1}\}} = [0]$ and $n_2 = 1$ during the first join. 
Substituting in Theorem \ref{thm:FirstJoinLp} we obtain the desired results. 
The rest of the $\kappa - 1$ edges are accounted for by edge firings 
(c.f. the discussion in \textsection\ref{Sec:TwoToOne}). 
Therefore, we need only $O(\kappa \cdot \tau^2)$ computations at time step $\tau$, and hence 
$O\left(\displaystyle \kappa \cdot \sum_{\tau = 1}^{n} \tau^2 = \kappa \cdot \frac{n (n+1) (2n + 1)}{6}\right)$, 
overall. As $\kappa << n$ in most practical cases, we have an order of magnitude lower average cost than 
that incurred by the standard methods. Further improvements follow from the parallelizability of our approach.        
Although we have not discussed it explicitly, it is evident that node and edge deletions can all be handled 
within this framework in the same way and at the same $O(n^2)$ cost per operation 
(c.f. the discussion in \textsection\ref{Sec:OneToTwo}). 
\begin{figure}[t]
\centerline{\begin{tabular}{ccc}
\scalebox{0.6}{\includegraphics[width = 70mm]{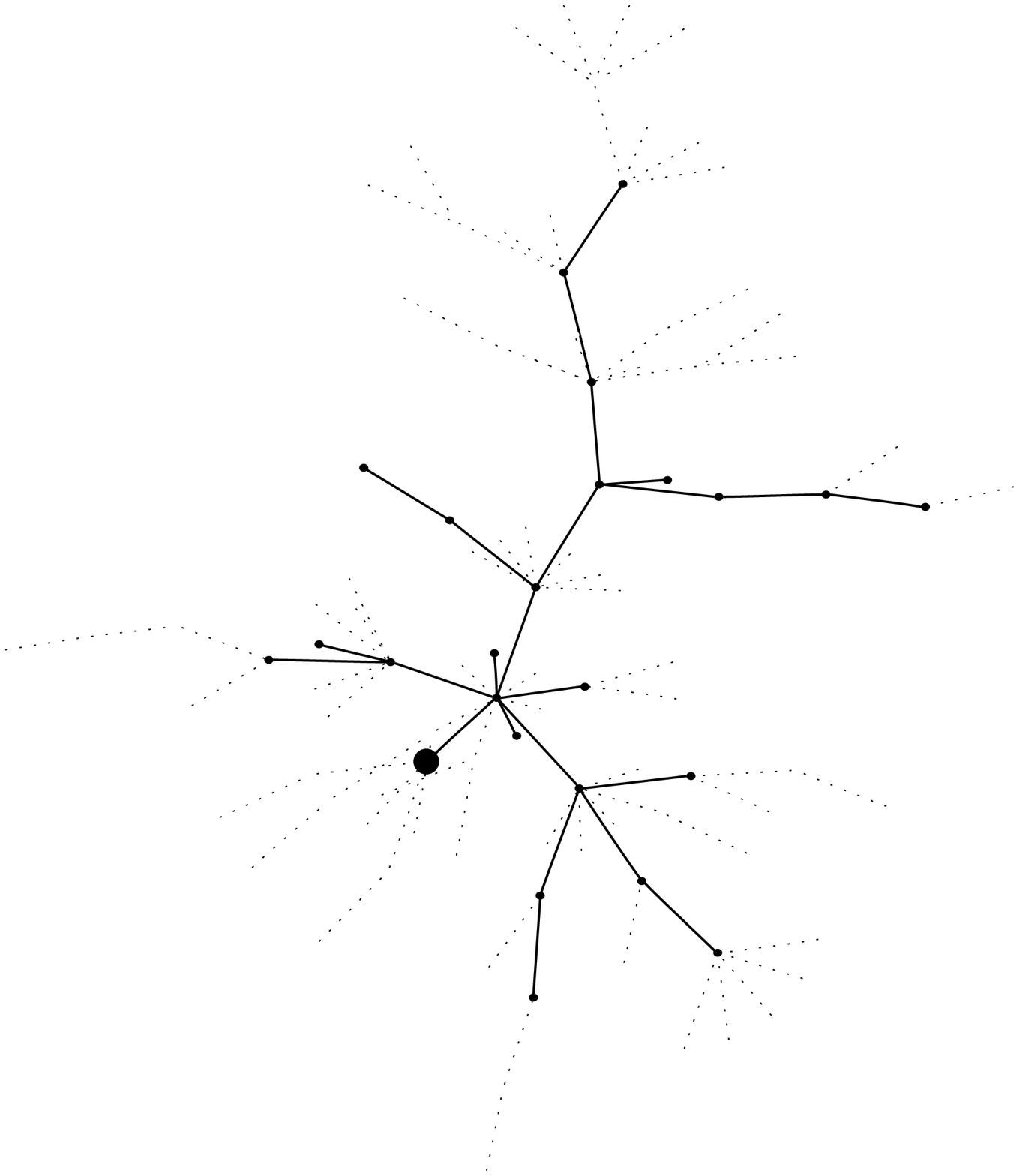}}
&
\scalebox{0.6}{\includegraphics[width = 70mm]{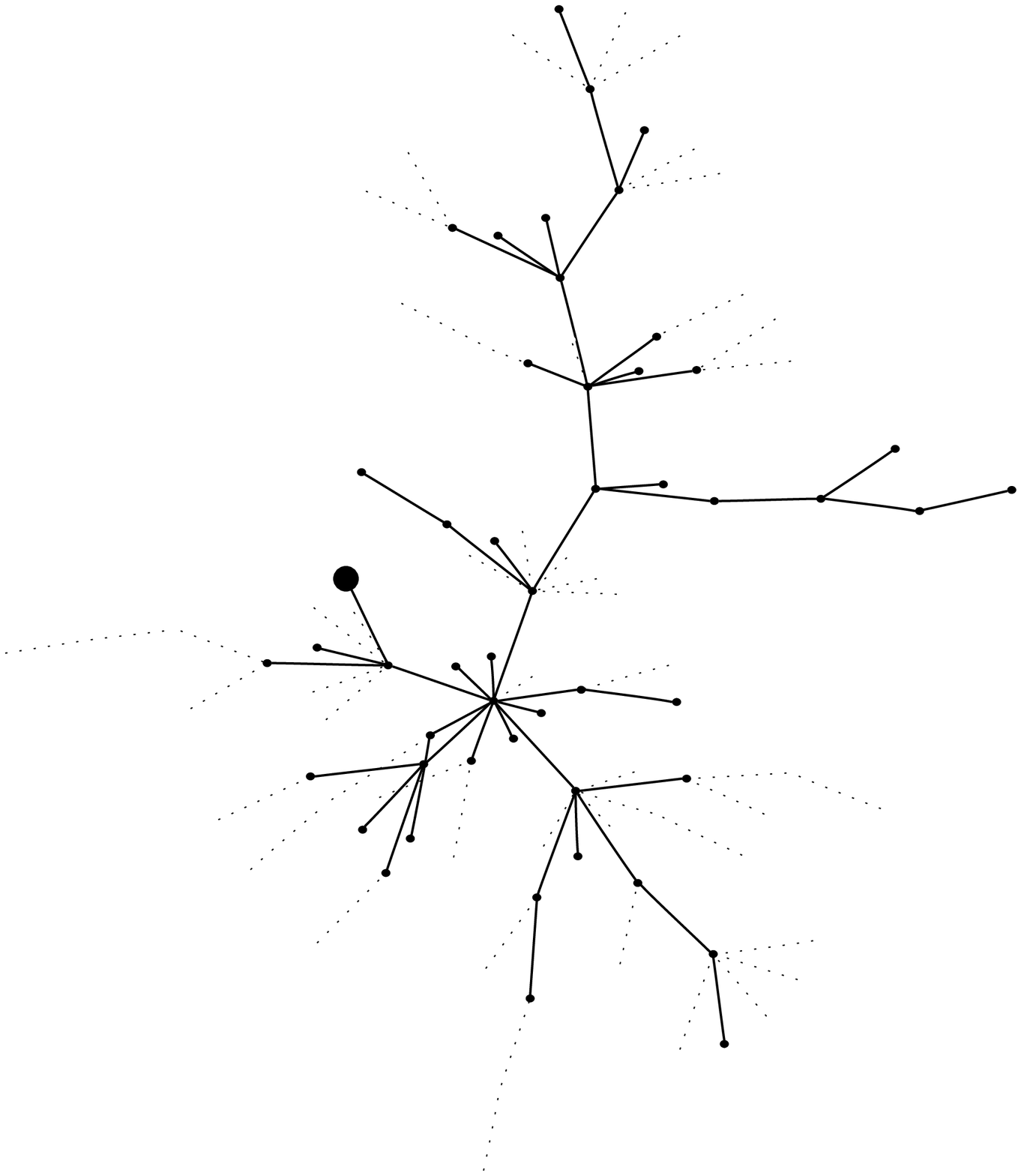}}
&
\scalebox{0.6}{\includegraphics[width = 70mm]{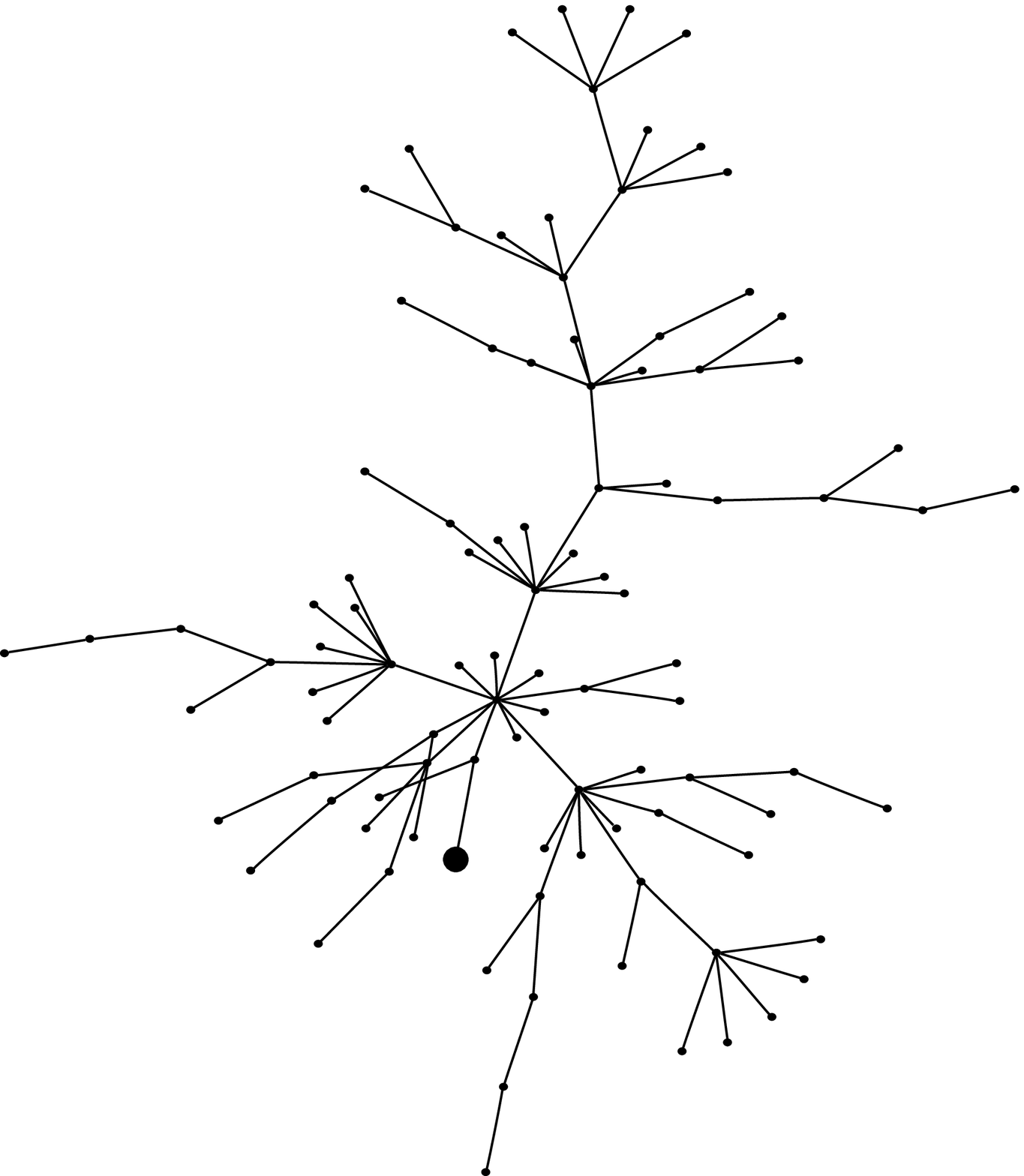}}
\\
(a) $\tau = 25$ & (b) $\tau = 50$ & $\tau = 100$
\end{tabular}}
\caption{Growing a tree by preferential attachment $(n = 100, ~\kappa = 1)$. 
The node $v_{\tau}$, being added to the tree at time step $\tau$, is emphasized (larger circle). 
Dotted edges at time steps $\tau = \{25, ~50\}$ are a visual aid representing  
edges that are yet to be added in the tree until the order-limit $(n = 100)$ is attained.}    
\label{fig:imageTreeByPA}
\end{figure}
\subsection{Large Real-World Networks: A Divide-And-Conquer Approach}
\label{SubSec:DivAndConq}
In order to compute $\bb L^+$ for an arbitrary graph $G(V, E)$, in a divide-and-conquer 
fashion, we need to first determine independent sub-graphs of $G$ in an efficient manner. 
Theoretically, an optimal divide step entails determining a {\em balanced} connected bi-partition 
$P(G_1, G_2)$ of the graph $G$ such that $|V(G_1)| \approx |V(G_2)|$ and $|E(G_1, G_2)|$, the 
number of edges violating the partition, is minimized. Such balanced bi-partitioning of the graph, 
if feasible, can then be repeated recursively until we obtain sub-graphs of relatively small orders. 
The solutions to these sub-problems can then be computed and the recursion unwinds to yield 
the final result, using our two-stage methodology in the respective conquer steps. Alas, 
computing such balanced bi-partitions, along with the condition of minimality of  $|E(G_1, G_2)|$, 
belongs to the class of {\em NP-Complete} problems \cite{Sen08}, and hence a polynomial time 
solution does not exist. We therefore need an efficient alternative to accomplish the task at hand. 
Partitioning of graphs to realize certain objectives has been studied extensively  
in diverse domains such as VLSI CAD \cite{AlpertKahng}, parallel computing, 
artificial intelligence and image processing  \cite{ShiMalik97}, 
and power systems modeling \cite{Sen08,Tiptipakorn01}. 
Perhaps, the most celebrated results in this class of problems are the spectral method 
\cite{Luxburg07} and the max-flow = min-cut \cite{FordFulkerson56}, both of which are 
computable in polynomial time \cite{Luxburg07, EdmondsKarp72}. 
Approximation algorithms for the balanced connected bi-partitions problem, for some 
special cases, have also been proposed \cite{Chlebikova96, Amir03}. Although useful 
in specific instances, such methods when used for the divide step may, in themselves, 
incur high computational costs thus undermining the gains of the conquer step. 
We need a simple methodology that works well on {\em real-world} networks. 
%
%
%
\begin{table}[t]
\begin{center}
\small
\begin{tabular}{||l|l|l|c||c|c|c|c|c||}
\hline
$G(V, E)$ & $n = |V(G)|$ & $m = |E(G)|$ & Leaves & Cut-off & \# Comp. & $|V(GCC)|$ & $|E(GCC)|$ & \# Cut-Edges 
\\
\hline\hline
{\em Epinions} &   75,888 & 405,740 & 35,763 & 4,429 & 30,376 & 37,924 & 61,482 & 102,452 
\\
\hline 
{\em SlashDot}  & 82,168 & 504, 230 & 28,499 & 7,012  & 36,311 & 41,084 & 62,225 & 164,719 
\\
\hline
\end{tabular}
\end{center}
\caption{Basic properties: Epinions and SlashDot networks.}
\label{Tab:RealOSN}
\end{table}
\begin{figure}[t]
\centerline{\begin{tabular}{ll}
\scalebox{0.6}{\includegraphics[width = 100mm]{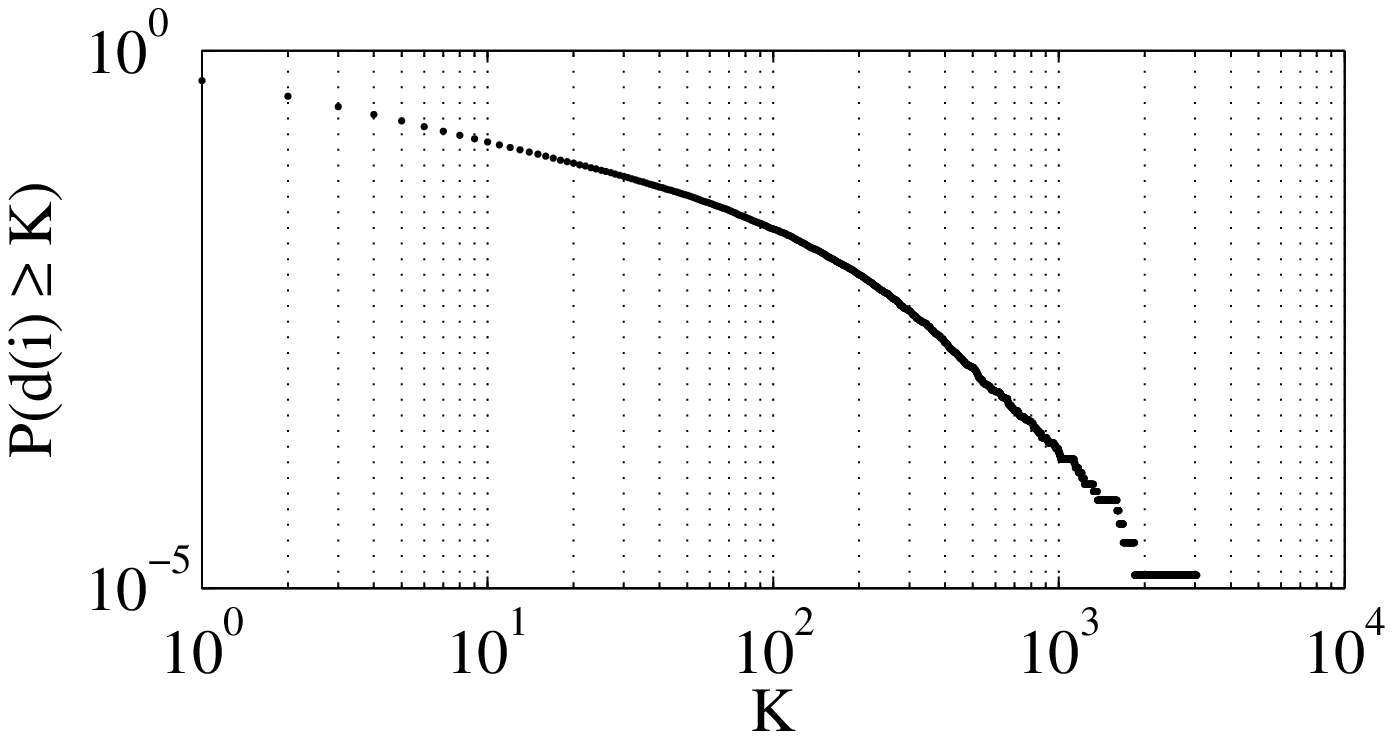}}
&
\scalebox{0.6}{\includegraphics[width = 100mm]{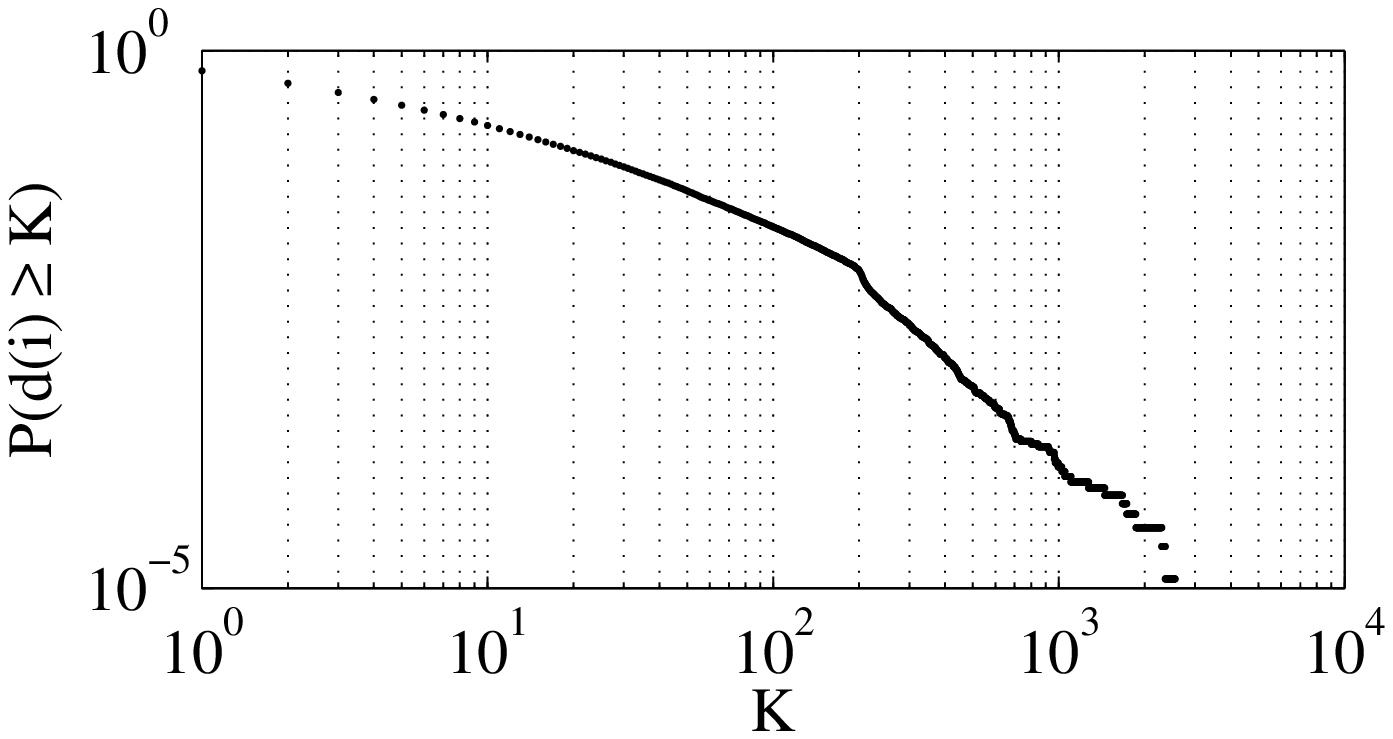}}
\\
(a) {\em Epinions}: Degree dist. & (b) {\em SlashDot}: Degree dist. 
\\
~~~\scalebox{0.55}{\includegraphics[width = 100mm]{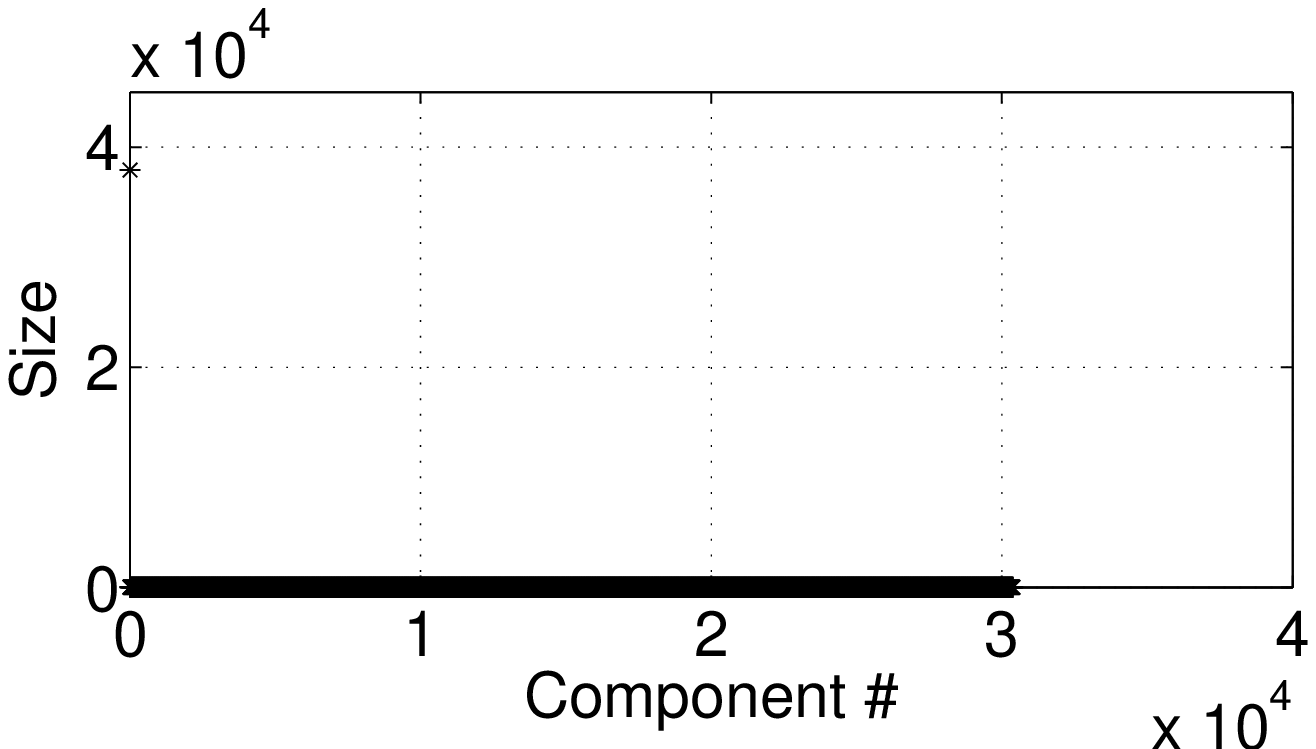}}
&
~~~\scalebox{0.55}{\includegraphics[width = 100mm]{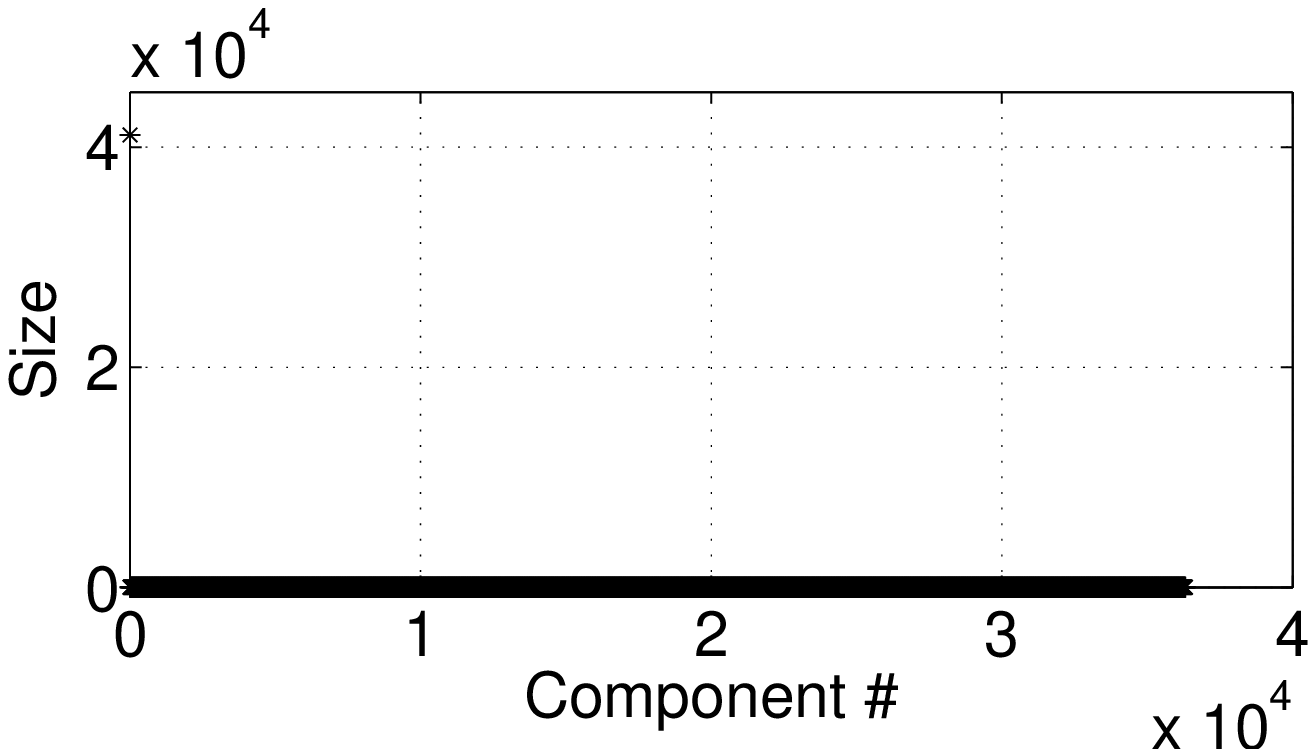}}
\\
(c) {\em Epinions}: Components at cut-off & (d) {\em SlashDot}: Components at cut-off
\end{tabular}}
\caption{Structural regress: Epinions and SlashDot Networks.}    
\label{fig:RealOSN}
\end{figure}

Real-world networks, and particularly online social networks, have been shown to have several 
interesting structural properties: edge sparsity, power-law scale-free degree distributions, 
existence of the so called {\em rich club connectivity} \cite{Barabasi00, Barabasi99, Barabasi01},  
small-world characteristics \cite{Watts98} with relatively small diameters ($O(log ~n)$). 
Collectively, these properties amount to a simple fact: the overall connectivity between arbitrary 
node pairs is dependent on higher degree nodes in the network. 
Based on these insights, we now study two real-world online social networks --- 
the {\em Epinions} and {\em SlashDot} networks \cite{snap} --- 
to attain our objective of a quick and easy divide step. Table \ref{Tab:RealOSN} gives some of the 
basic statistics about the two networks.\footnote{Although the networks originally have uni-directional 
and bi-directional links, we symmetrize the uni-directional edges to make the graphs undirected.} 
It is easy to see that the networks are sparse as $m = O(n) << O(n^2)$ in both cases.  
Moreover, note that a significant fraction of nodes in the graphs are leaf/pendant nodes, 
i.e. nodes of degree 1 ($\approx 47\%$ for Epinions and $\approx 34\%$ for {\em SlashDot}). 
From Fig. \ref{fig:RealOSN} (a-b), it is also evident that the node degrees indeed follow a heavy tail 
distribution in both cases.
Thus, there are many nodes of very small degree (e.g. leaves) and relatively fewer nodes of very high 
degrees in these networks. 
Therefore, in order to break the graph into smaller sub-graphs, we adopt an incremental regress 
methodology of deleting high degree nodes.  
Ordering the nodes in decreasing order by degree, we remove them one at a time. This process divides 
the set of nodes into three parts at each stage: 
\begin{enumerate}[a.]
\item \textbf{The Rich Club}: High degree nodes that have been deleted until that stage. 
\item \textbf{The Giant Connected Component} ($GCC$): The largest connected component at that stage.  
\item \textbf{Others}: All nodes that are neither in the rich club nor the $GCC$. 
\end{enumerate}
We repeat the regress, one node at a time, until the size of the $GCC$ is less than half the size 
of the original graph. We call this the cut-off point. We then retain the $GCC$ as one of our 
sub-graphs (one independent sub-problem) and re-combine all the non-$GCC$ nodes together 
with the rich club to obtain (possibly) multiple sub-graphs (other sub-problems). 
This concludes the divide step.                

Table \ref{Tab:RealOSN} shows the relevant statistics at the cut-off point for the two networks. 
Note that the cut-off point is attained at the expense of a relatively small number of high 
degree nodes ($\approx 5\%$ for {\em Epinions} and $\approx 8\%$ for {\em SlashDot}).  
Moreover, the number of nodes in the $GCC$ is indeed roughly half of the overall order, 
albeit the $GCC$ is surely sparser in terms of edge density than the overall network 
($|E(GCC)|/|V(GCC)| = 1.63$ vs. $|E(G)|/|V(G)| = 5.35$ for {\em Epinions} and 
$|E(GCC)|/|V(GCC)| = 1.51$ vs. $|E(G)|/|V(G)| = 6.13$ for {\em Slashdot}). 
Fig. \ref{fig:RealOSN} (c-d) shows the sizes (in terms of nodes) of all the connected components 
for the respective graphs at the cut-off point. It is easy to see that other than the $GCC$, 
the remaining components are of negligibly small orders. Recombining the non-$GCC$ components 
together (including the rich club) yields an interesting result. For the {\em Epinions} network, 
we obtain two sub-graphs of orders $37, 933$ and $31$ respectively while for the {\em Slashdot} 
network we obtain exactly one sub-graph of order $41,084$. This clearly demonstrates that 
our simple divide method, yields a roughly equal partitioning of the network --- 
and thus comparable sub-problems --- in terms of nodes. The pseudo-inverses of these sub-problems
can now be computed in parallel. 
Albeit, as in the case of all tradeoffs, this equitable split comes at a price of roughly $\kappa = O(n)$ 
edges that violate the cut (c.f. Table \ref{Tab:RealOSN}). This yields an $O(n^3)$ average cost for 
the {\em two-stage} process (c.f. \textsection\ref{Sec:TwoToOne}). However, given the element-wise 
parallelizability of our method, we obtain the pseudo-inverses in acceptable times of roughly 
$15$ minutes for the {\em Epinions} and $18$ minutes for the {\em SlashDot} networks.    
\

%
%

\section{Related Work} 
\label{Sec:RelWork}
The applications of the Moore-Penrose pseudo-inverse and the sub-matrix 
inverses of the Laplacian for a graph are multifarious. We discuss a few instances
here in summary. 
As alluded to earlier, $\bb L^+$ is used to compute effective resistance distances 
between the nodes of a graph \cite{KleinRandic93} as well as the one way hitting 
and commute times in random walks between node pairs in a graph. 
All these distances serve as measures of {\em multi-hop} 
dissimilarity between nodes and find applications in several graph mining contexts 
\cite{Chandra89, Fouss07}. Moreover, for every connected undirected graph there 
is an analogous {\em reversible} Markov chain, $\bb L^+$ finds use in the computation 
of relevant metrics (such as hitting time, cover time and mixing rates).      

$\bb L^+$ is a gram-matrix. Its eigen decomposition yields an $n-dimensional$ 
Euclidean embedding of the graph whereby each node in the graph is represented as
a point in that space. The general term $l^+_{xy}$ represents the inner product  
of the respective position vectors for the nodes $x$ and $y$ and thus $\bb L^+$ 
is a valid kernel for a graph.
This geometric interpretation has been used in collaborative recommendation systems  
\cite{Fouss07, FoussPirotteRendersSaerens05, FoussYenPirotteSaerens06}.  

In  \cite{ChebShamis97, ChebShamis98b} the elements of $\bb L^+$ have been given 
an elegant topological interpretation in terms of the {\em dense} spanning rooted forests 
of the graph.
Namely, the general term $l^+_{xy}$ represents the number of spanning rooted forests 
of the graph with exactly two trees in which the pair $(x, y)$ is in the same tree, rooted 
at $x$ (or $y$ by symmetry). 
Combining the geometric interpretation from 
\cite{Fouss07, FoussPirotteRendersSaerens05, FoussYenPirotteSaerens06} 
and the topological interpretation from \cite{ChebShamis97, ChebShamis98b}, 
the diagonal elements of $\bb L^+$ have been used as centrality indices 
for the nodes of a complex network  in \cite{RanjanZhang11a}. 
This centrality index, called topological centrality, reflects both the overall position 
as well as the overall connectedness of nodes in the network. 
Consequently, it is a measure of the robustness of node to random multiple 
failures in the network. 
By extension, $Tr(\bb L^+)$, also called the Kirchhoff index of a graph 
\cite{KleinRandic93, Xiao03}, is a global structural descriptor for the graph on a whole. 
This index is quite popular in the field of mathematical chemistry and is used to measure
overall molecular strength \cite{Palacios01b, Palacios10a, Palacios01a, Palacios10b}. 

The elements of sub-matrix inverses have analogous interpretations in terms of {\em unrooted}  
spanning forests of the graph \cite{Kirkland97}.  In \cite{Newman05}, the sub-matrix 
inverses have been used to compute the random-walk betweenness centrality, another useful 
index to characterize roles of nodes in a network.   
\

%
%

%
%

\section{Conclusion and Future Work}
\label{Sec:Conclusion}
In this work, we presented a divide-and-conquer based approach for computing the 
Moore-Penrose pseudo-inverse of the Laplacian ($\bb L^+$) for a simple, connected, 
undirected graph. Our method relies on an elegant interplay between the elements 
of $\bb L^+$ and the pairwise effective resistance distances in the graph. Exploiting 
this relationship, we derived closed form solutions that enable us to compute  
$\bb L^+$ in an incremental fashion. We also extended these results to analogous cases
for structural regress. Using dynamic networks and online social networks as examples, 
we demonstrated the efficacy of our method for computing the pseudo-inverse relatively 
faster than the standard methods. The insights from our work open up several interesting 
questions for future research. First and foremost, similar explorations can be done for 
the case of directed graphs (asymmetric relationships), where analogous 
distance functions --- such as the expected commute time in random walks --- are defined, 
albeit the Laplacians (more than one kind in literature) are no longer symmetric  \cite{Boley11}. 
Secondly, matrix-distance interplays of the kind exploited in this work, also exist for a general 
case of the so called {\em forest matrix} and its distance counterpart the {\em forest distance} 
\cite{Agaev00,ChebShamis97}, both for undirected and directed graphs. 
he results presented here should find natural extensions to the forest 
matrix and the forest metric, at least for the undirected case.  
Finally, our closed forms can be used in conjunction with several interesting approaches 
for sparse inverse computations \cite{Campbell95}, to further expedite the pseudo-inverse 
computation for large generalized graphs. All these motivate ample scope for future work.    
\

%
%

%
%
\begin{table}[t]
\begin{center}
\small
\begin{tabular}{||l|l|l||}
\hline
Operation & ~~~~~~~~~~~~~~~~~~~~~~~$\Omega$ & ~~~~~~~~~~~~~~~~~~~~~~~~~~~~~~~$\bb L^+$ \\
\hline\hline
%
%
~ 
& $x, y \in G_1: \Omega^{G_3}_{xy} = \Omega^{G_1}_{xy}$  
& $l^{+(1)}_{xy} - \frac{n_2 n_3 \left(l^{+(1)}_{xi} + l^{+(1)}_{iy}\right) - n_2^2 \left(l^{+(1)}_{ii} + l^{+(2)}_{jj} + \omega_{ij}\right)} {n_3^2}$ 
\\
First Join
&
$x, y \in G_2: \Omega^{G_3}_{xy} = \Omega^{G_2}_{xy}$ 
&
$l^{+(2)}_{xy} - \frac{n_1n_3 \left(l^{+(2)}_{xj} + l^{+(2)}_{jy}\right) - n_1^2 \left(l^{+(1)}_{ii} + l^{+(2)}_{jj} + \omega_{ij}\right)}{n_3^2}$ 
\\
~
&
$x \in G_1, y \in G_2: \Omega^{G_3}_{xy} = \Omega^{G_1}_{xi} + \omega_{ij} + \Omega^{G_1}_{jy}$ 
&
 ~~~~~~~~~~$\frac{n_3 \left( n_1l^{+(1)}_{xi} + n_2 l^{+(2)}_{jy}\right) - n_1 n_2 \left(l^{+(1)}_{ii} + l^{+(2)}_{jj} + \omega_{ij}\right)}{n_3^2}$ 
\\
\hline\hline
Edge firing 
&
$\Omega^{G_1}_{xy} 
- \frac{\left[\left(\Omega^{G_1}_{xj} - \Omega^{G_1}_{xi}\right) - \left(\Omega^{G_1}_{jy} - \Omega^{G_1}_{iy}\right)\right]^2}
{4(\omega_{ij} + \Omega^{G_1}_{ij})} $ 
&
$l^{+(1)}_{xy} 
- \frac{\left(l^{+(1)}_{xi} - l^{+(1)}_{xj}\right) \left(l^{+(1)}_{iy} - l^{+(1)}_{jy}\right)}{\omega_{ij} + \Omega^{G_1}_{ij}} 
$ 
\\
\hline\hline 
Non-bridge delete 
& $\Omega^{G_1}_{xy} 
+ \frac{\left[\left(\Omega^{G_1}_{xj} - \Omega^{G_1}_{xi}\right) 
- \left(\Omega^{G_1}_{jy} - \Omega^{G_1}_{iy}\right)\right]^2}
{4(\omega_{ij} - \Omega^{G_1}_{ij})}$ 
& $l^{+(1)}_{xy} + \frac{\left(l^{+(1)}_{xi} - l^{+(1)}_{xj}\right) \left(l^{+(1)}_{iy} - l^{+(1)}_{jy}\right)}{\omega_{ij} - \Omega^{G_1}_{ij}}$  
\\
\hline\hline 
Bridge delete
&
$x, y \in G_k: \Omega^{G_k}_{xy} = \Omega^{G_1}_{xy}$ 
&
$l^{+(1)}_{xy} - 
\frac{n_k \sum_{z \in G_k} \left(l^{+(1)}_{xz} 
+ l^{+(1)}_{zy}\right)
- \sum_{x \in G_k}\sum_{y \in G_k} l^{+(1)}_{xy}}
{n_k^2} 
$ 
\\
\hline
\end{tabular}
\end{center}
\caption{Summary of results: Atomic operations of the divide-and-conquer methodology.}
\label{Tab:SumOfRes}
\end{table}
%

%
%

\section{Acknowledgment}
\label{sec:Acknowledgment}
This research was supported in part by 
DTRA grant HDTRA1-09-1-0050 and NSF grants CNS-0905037, CNS-1017647, CNS-1017092
and IIS-0916750. 
\

%
%

%
%

\section{Appendix}
%
%
\subsection{Proof of Theorem \ref{thm:SubInvByLp}}
Given, $\bb L \in \Re^{n \times n}$, we note that $\bb L \cdot \bb 1 = \bb 0$ and $\bb 1' \cdot \bb L = \bb 0'$, 
where $\bb 1 \in \Re^n$ and $\bb 0 \in \Re^n$ are vectors of length $n$ containing all $1's$ 
and $0's$ respectively. 
From \cite{Boley11}, we have: 
\begin{equation}
\bb L(\{\overline n\},\{\overline n\})^{-1} =  [\bb I_{n - 1}, \bb{-v}] ~\bb L^+ 
\left[ \begin{array}{c}
\bb I_{n-1}\\
\bb{-u}'\\
 \end{array}\right] 
\end{equation}
where $\bb I_{n-1}$ is the identity matrix of dimension $(n - 1 \times n - 1)$ and 
$\bb u = \bb v = \bb 1 \in \Re^{n-1}$ are vectors of all $1's$ of length $n-1$. 
Expanding we obtain the following scalar form: 
\begin{equation}
[\bb L(\{\overline n\}, \{\overline n\})^{-1}]_{xy} = l^+_{xy} - l^+_{xn} - l^+_{ny} + l^+_{nn}  
\end{equation}

\noindent$\square$ 
\
%
%
\subsection{Proof of Lemma \ref{lem:SumOmegaTransitTriple}}
Given, $\bb L^+$ is symmetric and doubly centered and 
$\Omega_{xy} = l^+_{xx} + l^+_{yy} - l^+_{xy} - l^+_{yx}$, we have:  
\begin{eqnarray*}
\sum_{z=1}^n \Omega_{xz} + \Omega_{zy} - \Omega_{xy} 
&=& \sum_{z=1}^n [(l^+_{xx} + l^+_{zz} - 2l^+_{xz}) + (l^+_{zz} + l^+_{yy} - 2l^+_{zy}) - (l^+_{xx} + l^+_{yy} - 2l^+_{xy})]\\
&=& 2 \sum_{z=1}^n [l^+_{zz} + l^+_{xy}]\\
&=& 2n~  l^+_{xy} + 2 ~Tr(\bb L^+)   
\end{eqnarray*}
Rearranging terms, we get:
\begin{equation}
l^+_{xy} 
=  \frac{1}{2n}\left(\sum_{z=1}^{n} \Omega_{xz} + \Omega_{zy} - \Omega_{xy}\right) 
- \frac{1}{n} ~Tr(\bb L^+)   
\end{equation}
Substituting $\displaystyle Tr(\bb L^+) = \frac{1}{2n} \sum_{x = 1}^n \sum_{y = 1}^n \Omega_{xy}$, 
in the expression above, we obtain the proof. 

\noindent$\square$  
\
%
%
\subsection{Proof of Corollary \ref{cor:StarLp}}
Given, a star $S_p$ with node $1$ as root and nodes $\{2, 3, ..., p\}$ as leaves, we have: 
\begin{equation}
\forall x: 2 \leq x \leq p, ~~ 
\Omega^{S_p}_{1x} =  1 ~~~~~~ and ~~~~~~~
\forall x \neq y: 2 \leq x, y \leq p,  ~~ 
\Omega^{S_p}_{xy} = 2 
\end{equation}
\noindent Therefore, $$\sum_{x = 1}^p\sum_{y = 1}^p \Omega^{S_p}_{xy} = 2 (p - 1)^2$$  
Also, 
$$
\sum_{z = 1}^p \Omega^{S_p}_{1z} +   \Omega^{S_p}_{zp} -  \Omega^{S_p}_{1p} =  2 (p - 2)
 ~~~~~~ and ~~~~~~~
\forall x \neq y: 2 \leq x, y \leq p,  ~~ 
\sum_{z = 1}^p \Omega^{S_p}_{xz} +   \Omega^{S_p}_{zy} -  \Omega^{S_p}_{xy} =  2 (p - 3)
$$
Substituting for these values in Lemma \ref{lem:SumOmegaTransitTriple}, and noting that 
$\bb L^+_{S_p}$ is doubly-centered, we obtain the proof. 

\noindent $\square$  
%
%
\subsection{Proof of Corollary \ref{cor:CliqueLp}}
Given a clique $K_p$ of order $p$, $\forall x \neq y : 1 \leq x, y \leq p, \Omega_{xy} = \displaystyle \frac{2}{p}$ 
\cite{Biggs93}. Therefore,    
$$\sum_{x = 1}^p\sum_{y = 1}^p \Omega^{K_p}_{xy} = 2 (p - 1)$$

Also, 
$$ \forall x: 1 \leq x \leq p, ~~
\sum_{z = 1}^p \Omega^{K_p}_{xz} +   \Omega^{K_p}_{zx} -  \Omega^{K_p}_{xx} =   \frac{4(p -1)}{p}
$$

Substituting in Lemma \ref{lem:SumOmegaTransitTriple}, we obtain: $l^+_{xx} = \displaystyle \frac{p -1}{p^2}$ 
and, from the fact that $\bb L^+_{K_p}$ is doubly-centered, 
$\forall x \neq y: 1 \leq x, y \leq p, ~~l^+_{xy} = \displaystyle - \frac{l^+_{xx}}{p-1} =  -\frac{1}{p^2}$. 

\noindent $\square$  
%
%
%
\subsection{Proof of Theorem \ref{thm:FirstJoinLp}} 
We present the proofs for the following two cases: (a) $x, y \in V_1(G_1)$ 
and, (b) $x \in V_1(G_1)$ and $y \in V_2(G_2)$. 
\noindent It is obvious that the other two cases, viz. (c) $x, y \in V_2(G_2)$ 
and (d) $x \in V_2(G_2)$ and $y \in V_1(G_1)$, follow from symmetry.  
But first we must express $Tr(\bb L^+_{G_3})$ as a function of 
$(Tr(\bb L^+_{G_1}), Tr(\bb L^+_{G_2}))$, 
which is useful to us in both cases. 
%
%
\begin{mylem}
\label{lem:TraceByParts}
For two disjoint simple, connected, undirected graphs $G_1(V_1, E_1)$ and $G_2(V_2, E_2)$, 
let $G_3(V_3, E_3)$ be the graph resulting from the first join between $G_1$ and $G_2$ by 
means of introducing an edge $e_{ij}:  i \in V_1(G_1), j \in V_2(G_2)$. Then, 
\begin{equation}
\label{equ:TraceByParts}
Tr(\bb L^+_{G_3}) 
= Tr(\bb L^+_{G_1}) +Tr(\bb L^+_{G_2})
+ \frac{n_1n_2}{n_1+n_2} \left(l^{+(1)}_{ii} + l^{+(2)}_{jj} + \omega_{ij}\right)
\end{equation}
\end{mylem}
\subsubsection{Proof of Lemma \ref{lem:TraceByParts}}
For an arbitrary node $x \in V_1(G_1)$:   
\begin{eqnarray*}
\Omega^{G_3}_{xy} 
&=& \Omega^{G_1}_{xy}, ~~~~~~ ~~~~~~ ~~~~~~ ~~~~if ~y \in V_1(G_1) \\
&=&  \Omega^{G_1}_{xi} + \omega_{ij} +  \Omega^{G_2}_{jy}, ~~~~~if ~ y \in V_2(G_2)  
\end{eqnarray*}
Therefore, 
\begin{eqnarray*}
\sum_{y \in V_3(G_3)}\Omega^{G_3}_{xy} 
&=& 
\sum_{y \in V_1(G_1)} \Omega^{G_1}_{xy} 
+ \sum_{y \in V_2(G_2)}  \left(\Omega^{G_1}_{xi} + \omega_{ij} +  \Omega^{G_2}_{jy}\right)\\
&=& n_1 ~l^{+(1)}_{xx} + Tr(\bb L^+_{G_1}) 
+ n_2 \left(l^{+(1)}_{xx} + l^{+(1)}_{ii} - 2~l^{+(1)}_{xi}\right) + n_2 ~\omega_{ij} 
+ n_2 ~l^{+(2)}_{jj} + Tr(\bb L^+_{G_2}) 
\end{eqnarray*}
Summing up over all nodes $x \in V_1(G_1)$: 
\begin{eqnarray*}
\sum_{x \in V_1(G_1)}\sum_{y \in V_3(G_3)}\Omega^{G_3}_{xy} 
&=& 
(2~n_1 + n_2)~Tr(\bb L^+_{G_1})  + n_1 n_2 ~(l^{+(1)}_{ii} + l^{+(2)}_{jj} + \omega_{ij}) + n_1~Tr(\bb L^+_{G_2}) 
\end{eqnarray*}
By symmetry, 
\begin{eqnarray*}
\sum_{x \in V_2(G_2)}\sum_{y \in V_3(G_3)}\Omega^{G_3}_{xy} 
&=& 
(2~n_2 + n_1)~Tr(\bb L^+_{G_2})  + n_1 n_2 ~(l^{+(1)}_{ii} + l^{+(2)}_{jj} + \omega_{ij}) + n_2~Tr(\bb L^+_{G_1}) 
\end{eqnarray*}
Therefore, 
\begin{eqnarray*}
\sum_{x \in V_3(G_3)}\sum_{y \in V_3(G_3)}\Omega^{G_3}_{xy} 
&=& 
\sum_{x \in V_1(G_1)}\sum_{y \in V_3(G_3)}\Omega^{G_3}_{xy} 
+ 
\sum_{x \in V_2(G_2)}\sum_{y \in V_3(G_3)}\Omega^{G_3}_{xy} 
\\
&=&
2~(n_1 + n_2) \left(Tr(\bb L^+_{G_1}) +Tr(\bb L^+_{G_2})\right)
+ 2~n_1n_2 \left(l^{+(1)}_{ii} + l^{+(2)}_{jj} + \omega_{ij}\right)
\end{eqnarray*}
Now, substituting 
$\displaystyle Tr(\bb L^+_{G_3}) = \frac{1}{2n_3} \sum_{x \in V_3(G_3)}\sum_{y \in V_3(G_3)}\Omega^{G_3}_{xy}$, 
we obtain the result. 

\noindent $\square$ 
\subsubsection{Rest of the Proof of Theorem \ref{thm:FirstJoinLp}}
~
\newline
\textbf{Case a: $x, y \in V_1(G_1)$}
\newline\newline
From Lemma \ref{lem:SumOmegaTransitTriple}, we have: 
\begin{equation}
\label{equ:LpGeneralTermSameSide}
 l^{+(3)}_{xy} 
 = \frac{1}{2n_3}\left(\sum_{z \in V_3(G_3)}\Omega^{G_3}_{xz} + \Omega^{G_3}_{zy} - \Omega^{G_3}_{xy}\right) 
  - \frac{1}{n_3} Tr(\bb L^+_{G_3})  
\end{equation}
For the triangle inequality in the $RHS$ above:  
\begin{eqnarray*}
\sum_{z \in V_3(G_3)}\Omega^{G_3}_{xz} 
&=& \sum_{z \in V_1(G_1)}\Omega^{G_1}_{xz} 
+  \sum_{z \in V_2(G_2)}\left(\Omega^{G_1}_{xi} + \omega_{ij} + \Omega^{G_2}_{jz}\right) \\
&=& (n_1+n_2) ~l^{+(1)}_{xx} + Tr(\bb L^+_{G_1}) + n_2 ~l^{+(1)}_{ii} - 2n_2 ~l^{+(1)}_{xi} 
+ n_2 ~\omega_{ij}
+ n_2~l^{+(2)}_{jj} + Tr(\bb L^+_{G_2}) 
\end{eqnarray*}
By symmetry, 
\begin{eqnarray*}
\sum_{z \in V_3(G_3)}\Omega^{G_3}_{zy} 
&=& (n_1+n_2) ~l^{+(1)}_{yy} + Tr(\bb L^+_{G_1}) + n_2 ~l^{+(1)}_{ii} - 2n_2 ~l^{+(1)}_{yi} 
+ n_2 ~\omega_{ij}
+ n_2~l^{+(2)}_{jj} + Tr(\bb L^+_{G_2}) 
\end{eqnarray*}
Finally, for the last of the three terms: 
$$\displaystyle \sum_{z\in V_3(G_3)} \Omega^{G_3}_{xy} 
= (n_1 + n_2) \left(l^{+(1)}_{xx} + l^{+(1)}_{yy} - 2~l^{+(1)}_{xy}\right)
$$ 
\noindent Summing the three individual 
terms along with the value of $Tr(\bb L^+_{G_3})$ from ($\ref{equ:TraceByParts}$) 
and substituting the result in ($\ref{equ:LpGeneralTermSameSide}$), we obtain the proof. 
\newline\newline\newline
\textbf{Case b: $x \in V_1(G_1)$ and $y \in V_2(G_2)$}
\newline\newline
Once again, 
\begin{equation}
\label{equ:LpGeneralTermDiffSides}
 l^{+(3)}_{xy} 
 = \frac{1}{2n_3}\left(\sum_{z \in V_3(G_3)}\Omega^{G_3}_{xz} + \Omega^{G_3}_{zy} - \Omega^{G_3}_{xy}\right) 
  - \frac{1}{n_3}Tr(\bb L^+_{G_3})  
\end{equation}
For the triangle inequality in the $RHS$ above: 
\begin{eqnarray*}
\sum_{z \in V_3(G_3)} \Omega^{G_3}_{xz}  
&=&  
\sum_{z \in V_1(G_1)} \Omega^{G_1}_{xz} 
+ \sum_{z \in V_2(G_2)} \left(\Omega^{G_1}_{xi}  + \omega_{ij} + \Omega^{G_2}_{jz}\right)\\ 
&=& (n_1 + n_2)~ l^{+(1)}_{xx} + Tr(\bb L^+_{G_1}) 
+ n_2 ~l^{+(1)}_{ii} - 2n_2~  l^{+(1)}_{xi} + n_2 ~\omega_{ij} 
+  n_2 ~l^{+(2)}_{jj} + Tr(\bb L^+_{G_2})  
\end{eqnarray*}
Similarly, 
\begin{eqnarray*}
\sum_{z \in V_3(G_3)} \Omega^{G_3}_{zy}  
&=&  
\sum_{z \in V_1(G_1)} \left(\Omega^{G_2}_{yj}  + \omega_{ij} + \Omega^{G_1}_{iz}\right)  
+ \sum_{z \in V_2(G_2)} \Omega^{G_1}_{zy}\\ 
&=& (n_1 + n_2)~ l^{+(2)}_{yy} + Tr(\bb L^+_{G_2}) 
+ n_1 ~l^{+(2)}_{jj} - 2n_1~  l^{+(2)}_{jy} + n_1 ~\omega_{ij} 
+  n_1 ~l^{+(1)}_{ii} + Tr(\bb L^+_{G_1})  
\end{eqnarray*}
And finally, 
\begin{eqnarray*}
\sum_{z \in V_3(G_3)} \Omega^{G_3}_{xy} 
&=& 
\sum_{z \in V_3(G_3)} \Omega^{G_3}_{xi} + \omega_{ij} +  \Omega^{G_3}_{jy}\\
&=&
(n_1 + n_2) \left(l^{+(1)}_{xx} + l^{+(1)}_{ii} -2~l^{+(1)}_{xi} + \omega_{ij}  + l^{+(2)}_{yy} + l^{+(2)}_{jj} -2~l^{+(2)}_{jy}  \right)
\end{eqnarray*}
\noindent Summing the three individual 
terms along with the value of $Tr(\bb L^+_{G_3})$ from ($\ref{equ:TraceByParts}$) 
and substituting the result in ($\ref{equ:LpGeneralTermDiffSides}$), we obtain the proof. 

\noindent $\square$ 
\
%
%
\subsection{Proof of Theorem \ref{thm:EdgeFireLp}} 
We prove Theorem \ref{thm:EdgeFireLp} in two steps. 
First, in the following lemma, we provide a general result for 
a perturbation of a positive semi-definite matrix, 
which is then used to prove the overall theorem.     
%
%
\begin{mylem}
\label{lem:PertLpEdgeAddDel}
Given $V \in \Re^{n \times n}$ is a symmetric, positive semi-definite matrix 
and $X \in \Re^{n \times q}$ a perturbation matrix,  
such that $(I + \alpha X^T V^\dg X)$  has an inverse where $\alpha=\{1, -1\}$, 
and $V V^\dg X = X$, the following holds:   
\begin{equation}
(V + \alpha X X^T)^\dg 
= V^\dg - \alpha V^\dg X (I + \alpha X^T V^\dg X)^{-1} X^T V^\dg
\end{equation}
\end{mylem}
\subsubsection{Proof of Lemma \ref{lem:PertLpEdgeAddDel}:}
Observe that the positive semi-definiteness of $V$
guarantees that $I + X^T V^\dg X$  has an inverse.

Let $W = V^\dg - \alpha V^\dg X (I + \alpha X^T V^\dg X)^{-1} X^T V^\dg$.
Therefore,  
$${
\begin{array}{lll}
(V + \alpha X X^T) W
& = &
 VV^\dg - \alpha VV^\dg X (I + \alpha X^T V^\dg X)^{-1} X^T V^\dg \\
 &&  + ~  \alpha XX^TV^\dg - XX^TV^\dg X (I + \alpha X^T V^\dg X)^{-1} X^T V^\dg \\[1ex]
& = &
 VV^\dg - \alpha X (I + \alpha X^T V^\dg X)^{-1} X^T V^\dg \\
 &&  + ~  \alpha X \left[ I - \alpha X^TV^\dg X (I + \alpha X^T V^\dg X)^{-1} \right]X^T V^\dg\\[1ex]
& = &
 VV^\dg - \alpha X (I + \alpha X^T V^\dg X)^{-1} X^T V^\dg \\
 &&  + ~  \alpha X \left[ ((I + \alpha X^T V^\dg X) - \alpha X^TV^\dg X) (I + \alpha X^T V^\dg X)^{-1} \right]X^T V^\dg\\[1ex]
& = &
 VV^\dg - \alpha X (I + \alpha X^T V^\dg X)^{-1} X^T V^\dg \\
 &&  + ~  \alpha X \left[  (I + \alpha X^T V^\dg X)^{-1} \right]X^T V^\dg\\[1ex]
& = &
 VV^\dg - \alpha X (I + \alpha X^T V^\dg X)^{-1} X^T V^\dg \\
 &&  + ~  \alpha X \left[ (I + \alpha X^T V^\dg X)^{-1} \right]X^T V^\dg\\[1ex]
& = &
 VV^\dg 
\end{array}
}$$
From this identity, and the symmetry of $V, W ~\& ~(XX^T)$,
it follows easily that $W$ satisfies the four conditions required
for a Moore-Penrose pseudo-inverse.

\noindent $\square$ 
\subsubsection{Rest of the Proof of Theorem \ref{thm:EdgeFireLp}}
Note that the firing of the edge $e_{ij}$ in $G_1(V_1, E_1)$ to obtain 
$G_2(V_2, E_2)$, results in the following scalar relationships between the Laplacians 
of the two graphs: 
\begin{equation}
a.~~~ [\bb L_{G_2}]_{ij} = [\bb L_{G_2}]_{ji} = -\frac{1}{\omega_{ij}}, 
~~~~~b.~~~ [\bb L_{G_2}]_{ii} = [\bb L_{G_1}]_{ii} + \frac{1}{\omega_{ij}}, 
~~~~~c.~~~ [\bb L_{G_2}]_{jj} = [\bb L_{G_1}]_{jj} + \frac{1}{\omega_{ij}}
\end{equation}
For ease of exposition, we permute the rows and columns in 
$\bb L_{G_1}$ and  $\bb L_{G_2}$ in such a way that $i = 1$ and $j = 2$. 
The above perturbations can then be rewritten as: 
\begin{equation}
\bb L_{G_2} 
= ~~\bb L_{G_1} ~~~ + \frac{1}{\omega_{12}}~~~
\left[ \begin{array}{ccccc}
1 & -1 & 0 & ... & 0\\
-1 & 1 & 0 & ... & 0\\
0 & 0 & 0 & ... & 0\\ 
\vdots & \vdots & \vdots & \vdots & \vdots \\
0  & 0 & 0 & ... & 0 \end{array} \right] 
\end{equation}
\noindent   $\bb L_{G_2}$ is therefore a sum of $\bb L_{G_1}$, a real, symmetric, 
positive semi-definite matrix, and a $rank(1)$ perturbation matrix, referred to henceforth 
as $Y$. 
It is easy to see that for a simple, connected, undirected graph $G_1(V_1, E_1)$, 
$\bb L^+_{G_1}$  satisfies all the preconditions in 
Lemma \ref{lem:PertLpEdgeAddDel}.   
Substituting $V = \bb L_{G_1}$, 
$\alpha = 1$ and 
$X = \sqrt{Y} = X^T$,  in Lemma \ref{lem:PertLpEdgeAddDel}, we get:  
\begin{equation}
 \bb L_{G_2}^+ = ( \bb L_{G_1} + X X^T)^+ = \bb L_{G_1}^+ - \bb L_{G_1}^+ X~(I + X \bb L_{G_1}^+ X)^{-1}~ X \bb L_{G_1}^+
\end{equation}
\noindent All that remains now is to obtain the scalar form for the term:  
$\bb L_{G}^+ X~(I + X \bb L_{G}^+ X)^{-1}~ X \bb L_{G}^+$. 
\noindent Note: 
\begin{equation}
\bb L_{G_1}^+ X 
=
\frac{1}{\sqrt{2~\omega_{12}}} \left[ \begin{array}{cc|ccc}
l^{+(1)}_{11} - l^{+(1)}_{12} & -(l^{+(1)}_{11} - l^{+(1)}_{12})  & 0 & ... & 0 \\
l^{+(1)}_{21} - l^{+(1)}_{22} & -(l^{+(1)}_{21} - l^{+(1)}_{22})  & 0 & ... & 0 \\ 
\vdots & \vdots & \vdots & \vdots & \vdots \\
l^{+(1)}_{n-1,1} - l^{+(1)}_{n-1,2} & - (l^{+(1)}_{n-1,1} - l^{+(1)}_{n-1,2})  & 0 & ... & 0\\
l^{+(1)}_{n1} - l^{+(1)}_{n2} & -(l^{+(1)}_{n1} - l^{+(1)}_{n2})  & 0 & ... & 0\end{array} \right]  
\end{equation}
Similarly, 
\begin{equation}
X \bb L_{G_1}^+
=
\frac{1}{\sqrt{2~\omega_{12}}} \left[ \begin{array}{cccc}
l^{+(1)}_{11} - l^{+(1)}_{21} & l^{+(1)}_{12} - l^{+(1)}_{22}   & ... &  l^{+(1)}_{1n} - l^{+(1)}_{2n}\\
- (l^{+(1)}_{11} - l^{+(1)}_{21}) & -(l^{+(1)}_{12} - l^{+(1)}_{22}) & ... & -(l^{+(1)}_{1n} - l^{+(1)}_{2n})\\ 
\hline\\
 0 & 0  & ... & 0\\
\vdots & \vdots & \vdots & \vdots \\
0 & 0 & ... & 0\end{array} \right]  
\end{equation}
Or, 
\begin{equation}
X\bb L^+_{G_1} X =  \left[ \begin{array}{cc|ccc}
~\frac{\Omega^{G_1}_{12}}{2~\omega_{12}} & -\frac{\Omega^{G_1}_{12}}{2~\omega_{12}} & 0 & ... & 0\\
-\frac{\Omega^{G_1}_{12}}{2~\omega_{12}} &~ \frac{\Omega^{G_1}_{12}}{2~\omega_{12}} & 0 & ... & 0\\
\hline
0 & 0 & 0 & ... & 0\\ 
\vdots & \vdots & \vdots & \vdots & \vdots \\
0  & 0 & 0 & ... & 0 \end{array} \right] 
\end{equation}
\noindent where $\Omega^{G_1}_{12} = l^{+(1)}_{11} + l^{+(1)}_{22} - l^{+(1)}_{12} - l^{+(1)}_{21}$,  
which yields:
\begin{equation}
(I + X\bb L^+_{G} X)^{-1} 
=  \left[ \begin{array}{cc|ccc}
1 + \frac{\Omega^{G_1}_{12}}{2~\omega_{12}} & -\frac{\Omega^{G_1}_{12}}{2~\omega_{12}} & 0 &... & 0\\
-\frac{\Omega^{G_1}_{12}}{2~\omega_{12}} & 1 + \frac{\Omega^{G_1}_{12}}{2~\omega_{12}} & 0 &... & 0\\
\hline
0 & 0 & ~ & ~ & ~\\ 
\vdots & \vdots & ~ & I_{n-2, n-2} & ~ \\
0  & 0 & ~ & ~ & ~ \end{array} \right] ^{-1} 
= 
 \left[ \begin{array}{cc|ccc}
\frac{2\omega_{12} + \Omega^{G_1}_{12}}{2 (\omega_{12} + \Omega^{G_1}_{12})} & \frac{\Omega^{G_1}_{12}}{2 (\omega_{12} + \Omega^{G_1}_{12})} & 0 &... & 0\\
\frac{\Omega^{G_1}_{12}}{2 (\omega_{12} + \Omega^{G_1}_{12})} & \frac{2\omega_{12} + \Omega^{G_1}_{12}}{2 (\omega_{12} + \Omega^{G_1}_{12})} & 0 &... & 0\\
\hline
0 & 0 & ~ & ~ & ~\\ 
\vdots & \vdots & ~ & I_{n-2, n-2} & ~ \\
0  & 0 & ~ & ~ & ~ \end{array} \right] 
\end{equation}
Multiplying on the left and right sides of the $RHS$ above with $\bb L^+_{G_1} X$ 
and $X\bb L^+_{G_1}$ respectively, the following scalar form is obtained: 
\begin{equation}
l^{+(2)}_{xy} 
= 
l^{+(1)}_{xy} - \frac{(l^{+(1)}_{x1} - l^{+(1)}_{x2}) (l^{+(1)}_{1y} - l^{+(1)}_{2y})}{\omega_{12} + \Omega^{G_1}_{12}} 
\end{equation}
\noindent Substituting $i = 1$ and $j = 2$ back into the equation above, we obtain the proof. 

\noindent $\square$ 
\
%
%
\subsection{Proof of Corollary \ref{cor:EdgeFireOmega}}
Noting $\Omega^{G_2}_{xy} = l^{+(2)}_{xx} + l^{+(2)}_{yy} - l^{+(2)}_{xy} - l^{+(2)}_{yx}$ 
and substituting into the result of Theorem \ref{thm:EdgeFireLp}, we obtain the proof. 

\noindent $\square$ 
%
%
\subsection{Proof of Theorem \ref{thm:NonBridgeEdgeDelLp}}
Deleting a non-bridge edge $e_{ij} \in E_1(G_1)$ from $G_1(V_1, E_1)$ to obtain 
$G_2(V_2, E_2)$, results in the following scalar relationships between the Laplacians 
of the two graphs: 
\begin{equation}
a.~~~ [\bb L_{G_2}]_{ij} = [\bb L_{G_2}]_{ji} = \frac{1}{\omega_{ij}}, 
~~~~~b.~~~ [\bb L_{G_2}]_{ii} = [\bb L_{G_1}]_{ii} - \frac{1}{\omega_{ij}}, 
~~~~~c.~~~ [\bb L_{G_2}]_{jj} = [\bb L_{G_1}]_{jj} - \frac{1}{\omega_{ij}}
\end{equation}
Once again, for convenience, we rearrange the rows and columns of $\bb L_{G_1}$ and 
$\bb L_{G_2}$ in such a way that $i = 1$ and $j = 2$. Thus, 
\begin{equation}
\bb L_{G_2} 
= ~~\bb L_{G_1} ~~~ - \frac{1}{\omega_{12}}~~~
\left[ \begin{array}{ccccc}
1 & -1 & 0 & ... & 0\\
-1 & 1 & 0 & ... & 0\\
0 & 0 & 0 & ... & 0\\ 
\vdots & \vdots & \vdots & \vdots & \vdots \\
0  & 0 & 0 & ... & 0 \end{array} \right] 
\end{equation}
The rest of the proof follows as in the proof of Theorem \ref{thm:EdgeFireLp}, with the following modification.  
Substituting $V = \bb L_{G_1}$, $\alpha = -1$ and 
$X = \sqrt{Y} = X^T$,  in Lemma \ref{lem:PertLpEdgeAddDel},  
we get:  
\begin{equation}
 \bb L_{G_2}^+ = ( \bb L_{G_1} - X X^T)^+ = \bb L_{G_1}^+ + \bb L_{G_1}^+ X~(I + X \bb L_{G_1}^+ X)^{-1}~ X \bb L_{G_1}^+
\end{equation}
which yields the following scalar form:
\begin{equation}
l^{+(2)}_{xy} 
= 
l^{+(1)}_{xy} + \frac{(l^{+(1)}_{x1} - l^{+(1)}_{x2}) (l^{+(1)}_{1y} - l^{+(1)}_{2y})}{\omega_{12} - \Omega^{G_1}_{12}} 
\end{equation}
\noindent Substituting $i = 1$ and $j = 2$ back into the equation above, we obtain the proof. 

\noindent $\square$ 
%
%
\subsection{Proof of Corollary \ref{cor:NonBridgeEdgeDelOmega}}
Noting $\Omega^{G_2}_{xy} = l^{+(2)}_{xx} + l^{+(2)}_{yy} - l^{+(2)}_{xy} - l^{+(2)}_{yx}$ 
and substituting into the result of Theorem \ref{thm:NonBridgeEdgeDelLp}, we obtain the proof. 

\noindent $\square$ 
%
%
\subsection{Proof of Theorem \ref{thm:BridgeEdgeDelLp}}
We present the proof for the case: $x, y \in V_2(G_2)$ as the other case follows by symmetry.  
Once again, we need a lemma to determine $Tr(\bb L^+_{G_2})$ in terms of the 
elements of $\bb L^+_{G_1}$. 
%
%
\begin{mylem}
\label{lem:TraceDelBridge}
Let $G_1(V_1, E_1)$  be a simple, connected, unweighted graph with a bridge edge 
$e_{ij}:  i \in E_1(G_1)$ which upon deletion produces two disjoint simple graphs 
$G_2(V_2, E_2)$ and $G_3(V_3, E_3)$. Then, 
\begin{equation}
\label{equ:TraceDelBridge}
Tr(\bb L^+_{G_2}) 
= \sum_{x \in V_2(G_2)} l^{+(1)}_{xx}
- \frac{1}{n_2} \sum_{x \in V_2(G_2)}\sum_{y \in V_2(G_2)}  l^{+(1)}_{xy}
\end{equation}
\end{mylem}
\subsubsection{Proof of Lemma \ref{lem:TraceDelBridge}}
The proof follows simply by observing $\Omega_{xy}^{G_2} = \Omega_{xy}^{G_3}$, 
$\forall(x, y) \in V_2(G_2) \times V_2(G_2)$ and substituting values in terms of the 
elements of $\bb L^+_{G_1}$.   

\noindent $\square$ 
\subsubsection{Rest of the Proof of Theorem \ref{thm:BridgeEdgeDelLp}}
Follows similarly from the triangle inequality in Lemma \ref{lem:SumOmegaTransitTriple},  
by confining to node pairs $(x, y) \in V_2(G_2) \times V_2(G_2)$, and then substituting 
the result of Lemma \ref{lem:TraceDelBridge} and other relevant 
effective resistance values in terms of $\bb L^+_{G_1}$.  

\noindent $\square$ 
\

%
%

%
%

\bibliography{allrefs}
\bibliographystyle{abbrv}

%
%

\end{document}